\documentclass[graphicx,superscriptaddress]{revtex4-1}

\usepackage{graphics}
\usepackage{dcolumn}
\usepackage{bm}
\usepackage{float}
\usepackage[caption = false]{subfig}
\usepackage{amsmath}
\usepackage{amsfonts}
\usepackage{amssymb}
\usepackage{mathtools}
\usepackage{enumitem}
\usepackage{xcolor,colortbl}
\usepackage{xifthen}
\usepackage{csvsimple}

\newcommand{\bk}{\mathbf{k}}

\newcommand{\bq}{\mathbf{q}}

\newcommand{\br}{\mathbf{r}}

\newcommand{\bS}{\mathbf{S}}

\newcommand{\rhq}{\rho_{\bq}}

\begin{document}
\title{Multi-component generalized mode-coupling theory: Predicting dynamics from structure in glassy mixtures}


\author{Simone Ciarella}
\thanks{These authors contributed equally to this work}
\affiliation{Soft Matter and Biological Physics, Department of Applied Physics, Eindhoven University of Technology,  P.O.~Box 513, 5600 MB Eindhoven, The Netherlands}
\affiliation{Laboratoire de Physique de l’Ecole Normale Supérieure, ENS, Université PSL, CNRS, Sorbonne Université, Université de Paris, F-75005 Paris, France}
\author{Chengjie Luo}
\thanks{These authors contributed equally to this work}
\affiliation{Soft Matter and Biological Physics, Department of Applied Physics, Eindhoven University of Technology,  P.O.~Box 513, 5600 MB Eindhoven, The Netherlands}
\author{Vincent E.~Debets}
\affiliation{Soft Matter and Biological Physics, Department of Applied Physics, Eindhoven University of Technology,  P.O.~Box 513, 5600 MB Eindhoven, The Netherlands}
\author{Liesbeth M.~C.~Janssen}
\affiliation{Soft Matter and Biological Physics, Department of Applied Physics, Eindhoven University of Technology,  P.O.~Box 513, 5600 MB Eindhoven, The Netherlands}

%

\begin{abstract}
The emergence of glassy dynamics and the glass transition in dense disordered systems is still not fully understood theoretically. 
Mode-coupling theory (MCT) has shown to be effective in describing some of the non-trivial features of glass formation, but it cannot explain the full glassy phenomenology due to the strong approximations on which it is based.
Generalized mode-coupling theory (GMCT) is a hierarchical extension of the theory, which is able to outclass MCT by carefully describing the dynamics of higher order correlations in its generalized framework.
Unfortunately, the theory has so far only been developed for single component systems and as a result works poorly for highly polydisperse materials. 
In this paper, we solve this problem by developing GMCT for multi-component systems. We use it to predict the glassy dynamics of the binary Kob-Andersen Lennard-Jones mixture, as well as its purely repulsive Weeks-Chandler-Andersen analogue.
Our results show that each additional level of the GMCT hierarchy gradually improves the predictive power of GMCT beyond its previous limit. 
This implies that our theory is able to harvest more information from the static correlations, thus being able to better understand the role of attraction in supercooled liquids from a first-principles perspective.
\end{abstract}

%
%
\maketitle
%
\section{Introduction}
Understanding how supercooled liquids become rigid and turn into amorphous solids is still one of the major challenges in condensed matter physics~\cite{Anderson1995,Lubchenko2007,Berthier2011,Langer2014}.
This so-called glass transition is not a transition in the thermodynamic sense~\cite{Biroli2013}, but it is defined by the dramatic increase in viscosity (or
relaxation time) upon only a relatively slight change in thermodynamic control
parameters, e.g., temperature or density~\cite{debenedetti2001supercooled,Ediger1996}. 
This sudden and highly non-linear dynamic response is accompanied by only subtle changes in the microscopic structure,  rendering it difficult to identify the main physical mechanisms underlying the glass transition~\cite{Liesbeth2018front,Angell1995,binder2011glassy,Tarjus2011,Cavagna2009}.

In principle, it is widely accepted that the dynamics of each material is ultimately related to its structure~\cite{Royall2015}, and numerous theories have also aimed to exploit this idea to describe the glass transition~\cite{Biroli2012,Xia1999,Dell2015,Tarjus2005a,Sausset2008,Gotze1992,Reichman2005}.
Among these, mode-coupling theory (MCT) stands out as one of the few theories which is entirely based on first principles~\cite{Gotze1992,Reichman2005,Liesbeth2018front,gotze2008complex}.
This theory seeks to predict the full microscopic relaxation dynamics of a
glass-forming system (as a function of time, temperature, density, and
wavenumber $k$) based solely on knowledge of simple structural material
properties, such as the static structure factor $S(k)$.
Although the theory is often not fully quantitatively accurate, MCT has enjoyed success in predicting e.g.\ multistep relaxation patterns and universal scaling laws in the dynamics,
stretched exponentials and growing dynamical length scales~\cite{Gotze1999,Kob2002,Biroli2006}. Furthermore, the theory offers a qualitative and physically intuitive account of glass formation in terms of the so-called cage effect~\cite{Kob2002}. 
The (most severe) limitation of MCT lies, however, in its assumption of Gaussian correlations which causes 
noticeable discrepancy between the theory and experiments.

So far, promising methodical MCT correction efforts have been put forward for single-component systems
--or equivalently systems with a small degree of polydispersity-- using higher order
field-theoretic loop expansions~\cite{SzamelPRL2003,Wu2005,Janssen2015a,Mayer2006,JanssenPRE2014,Janssen2016a}. Results show that such an expansion 
can indeed be accomplished, producing a novel, hierarchical first-principles theory known as generalized MCT (GMCT). By systematically developing the hierarchical equations, GMCT has already proven to be capable of predicting
the microscopic dynamics of glassy materials with near-quantitative accuracy in
the low to moderately supercooled regime~\cite{Janssen2015a}.
Similar to MCT, this generalized framework also
 requires only static
structure as input and has no free parameters. 
Furthermore, GMCT also provides predictive insights into regimes of previously 
inaccessible dynamics for single-component glassy systems~\cite{Luo2020_1,Luo2020_2,Janssen2015a}, and notably preserves the celebrated scaling laws of standard MCT~\cite{Luo2020_1,Luo2020_2}. 

Unfortunately, single-component systems are a poor representation of most studied glasses, which are typically polydisperse and as a result exhibit different overall dynamics compared to monodisperse systems~\cite{Klochko2020,Voigtmann_2011}. In experiments polydispersity is for the most part
inevitable, while in computer simulations it is often added to hinder and prevent crystal formation~\cite{Kob1994}.
Noticeably, techniques such as Monte Carlo (MC) swaps 
capitalize onto polydispersity to achieve faster relaxation dynamics and
explore the free energy landscape in uncommon ways~\cite{Ninarello2017a}.
Binary polydisperse systems are the simplest generalization of single-component systems in this direction.
They add only a supplementary component to the mix and are able to
retain most of the advantages of a polydisperse systems while adding the least possible complexity. 

In this paper, we set out to extend the GMCT framework to systems with an arbitrary number of species, and seek to apply the newly developed framework to describe arguably the 
most famous and simple examples of binary glassy systems: the Kob-Andersen binary Lennard-Jones (KABLJ) mixture~\cite{Kob1994} and its 
purely repulsive version based on the Weeks-Chandler-Andersen (WCA) potential~\cite{wca1971}. These systems have been extensively studied in the past and comparisons with MCT have identified the existence of a region where
MCT is already in its non-ergodic phase, thus predicting a glass, while simulations at the same temperature and density indicate a supercooled liquid phase~\cite{Berthier2010,Berthier2011a,Landes2020,Dell2015,Banerjee2014,Coslovich2012}. In other words, a discrepancy still exists between the simulations and MCT, even when MCT is extended to multi-component systems~\cite{Nagele1999a,Gotze2003,Voigtmann2003,Weysser2010,Ruscher2020}. 
Here, after demonstrating the ability of GMCT to systematically tackle this discrepancy, 
we will also address a fundamental question regarding the simplest ingredients required to describe the dynamics of binary supercooled liquids.
Due to the fact that standard MCT--which is based solely on $S(k)$--fails in predicting the precise location of the glass transition, it could be concluded that higher order correlations are required~\cite{Coslovich2012,Berthier2010,Berthier2011a}.
However, GMCT can circumvent this failure. It again uses only $S(k)$ as input, but in a more refined set of equations which can translate structural properties into dynamical ones in a more accurate manner. 
Applying our multi-component GMCT to both mixtures, we will conclude that each level of the GMCT hierarchy provides a significant improvement in the prediction of the glass transition, finally conjecturing that the infinite hierarchy might be able to accurately predict the glassy dynamics from $S(k)$ only.

\section{Multi-component GMCT}
\label{sec:mgmct}

Multi-component GMCT is derived starting from the Mori-Zwanzig approach~\cite{zwanzig60,mori65} to predict the dynamics of density correlations, similarly to standard MCT.
However, while standard MCT amounts to a single integro-differential equation closed by a factorization approximation~\cite{Liesbeth2018front,gotze2008complex,Gotze1992,Reichman2005,Voigtmann2003}, GMCT is a hierarchy of nested integro-differential equations~\cite{Janssen2015a}.
Each level of this hierarchy represents an MCT-like dynamical equation for a higher order, multi-point density correlation function, which we recursively solve and use to predict the dynamics of the correlations at the lower levels.
Since a solution of this hierarchy is well defined for any self-consistent closure or truncation of the hierarchy~\cite{Biezemans2020}, we can formally continue the GMCT scheme up to arbitrary order to include as many higher-order correlations as desired.

In the Supplementary Information, we report the full derivation of multi-component GMCT for an $M$-component mixture.
To summarize it here, we introduce the main objects of the theory, i.e.\ the species-dependent density modes:
\begin{equation}
    \rhq^{\alpha}(t)=\frac{1}{\sqrt{N}}\sum_{i=1}^{N_{\alpha}}e^{-i\bq\cdot\br_i(t)}~.
\end{equation}
Here $\bq$ is a wavevector of length $q=|\bq|$, $t$ is the time, the index $\alpha$ represents one of the $M$ species, 
$N_{\alpha}$ is the number of particles that belong to the species $\alpha$, and $N=\sum_{\alpha=1}^{M}N_{\alpha}$ is the total number of particles in the system .
To simplify our equations we introduce the notation that $\{x_i\}$ is a list $x_1,...,x_n$ and $\{x_i\}/x_j$ is the same ordered list $\{x_i\}$ where the specific element $x_j$ has been removed.
In solving multi-component GMCT we are interested in determining the dynamical equation of the density correlations of order $n$. These dynamical correlations are defined as:
\begin{eqnarray}
F^{(n)}_{\{\alpha_i\};\{\beta_i\}}(\{k_i\},t)=\langle\rho^{\alpha_1}_{-k_1}...\rho^{\alpha_n}_{-k_n}\rho^{\beta_1}_{k_1}(t)...\rho^{\beta_n}_{k_n}(t)\rangle~,
\end{eqnarray}
where the angular brackets denote an ensemble average.
In particular, when the order $n=1$, the correlation corresponds to the intermediate scattering function.
The only required input of the theory (aside from temperature and density) is the set of wavevector-dependent static correlations $S^{(n)}$,
\begin{eqnarray}
&&S^{(n)}_{\{\alpha_i\};\{\beta_i\}}(\{k_i\})=F^{(n)}_{\{\alpha_i\};\{\beta_i\}}(\{k_i\},t=0) \approx
\prod_{i=1}^{n}S^{\alpha_i\beta_i}({k_i})~,
\label{eq:Sn}
\end{eqnarray}
which defines the full microstructure of the system at any given temperature and density. Here we factorize these static correlations as products of the two-point correlation $S^{\alpha\beta}({k})\equiv S^{(1)}_{\alpha;\beta}(k)$, which is also known as the static structure factor.
This means that all the predictions of the theory are based on the structure factor only. 
As such, we will conclude later that all the predicted differences in dynamics among the two supercooled liquid systems are already encoded in their static structure factors. 

Within our multi-component GMCT hierarchy, each dynamical correlation function $F^{(n)}_{\{\alpha_i\};\{\beta_i\}}(\{k_i\},t)$ obeys the following equation of motion:
\begin{align}
\ddot{F}^{(n)}_{\{\alpha_i\};\{\beta_i\}}(\{k_i\},t)&+\mu\dot{F}^{(n)}_{\{\alpha_i\};\{\beta_i\}}(\{k_i\},t)+F^{(n)}_{\{\alpha_i\};\{\beta_i\}}(\{k_i\},t)(S^{(n)})^{-1}_{\{\alpha_i\};\{\beta_i\}}(\{k_i\})J^{(n)}_{\{\alpha_i\};\{\beta_i\}}(\{k_i\})
\nonumber\\&
+\int_0^t d\tau \dot{F}^{(n)}_{\{\alpha_i\};\{\beta_i\}}(\{k_i\},t-\tau) (J^{(n)})^{-1}_{\{\alpha_i\};\{\beta_i\}}(\{k_i\})K^{(n)}_{\{\alpha_i\};\{\beta_i\}}(\{k_i\},\tau)=0~,
\label{eq:order-n}
\end{align}
where $\mu$ is an effective friction coefficient. In the above notation, 
multiplications of the species-dependent quantities are multiplications of matrices for a given $\{k_i\}$. The matrices $J$ are static elements defined as
\begin{align}
J^{(n)}_{\{\alpha_i\};\{\beta_i\}}(\{k_i\})&=
    \langle\frac{d}{dt}\left[\rho^{\alpha_1}_{k_1}...\rho^{\alpha_n}_{k_n}\right]|\frac{d}{dt}\left[\rho^{\beta_1}_{k_1}...\rho^{\beta_n}_{k_n}\right]\rangle /N^n
    \nonumber\\&
\approx
\sum_{i=1}^{n}\delta_{\alpha_i,\beta_i}\frac{k_BTx_{\alpha_i}k_i^2}{m_{\alpha_i}S^{\alpha_i\beta_i}}({k_i})\prod_{j=1}^{n}S^{\alpha_j\beta_j}(k_j)~.
\end{align}
with $T$ the temperature, $m_\alpha$ the particle mass of type $\alpha$, and $x_\alpha=N_\alpha/N$ their number ratio.
Each level $n$ of the hierarchy is connected to the next via the memory term
\begin{align}
&K^{(n)}_{\{\alpha_i\};\{\beta_i\}}(\{k_i\},\tau)=
\nonumber\\&
=\frac{\rho}{2}\sum_{\mu'\nu'}\sum_{\mu\nu}\int \frac{d\mathbf{q}}{(2\pi)^3}\cdot
\sum_{j=1}^{n}\frac{k_BT}{m_{\alpha_j}}\mathcal{V}_{\mu'\nu'\alpha_j}(\mathbf{q,k_j-q,k_j})F^{(n+1)}_{\mu',\nu',\{\alpha_i\}/\alpha_j;\mu,\nu,\{\beta_i\}/\beta_j}(\mathbf{q,k_j-q},\{k_i\}/k_j,\tau) 
\nonumber\\
&\qquad\qquad\qquad\qquad\qquad\qquad\cdot
\mathcal{V}_{\mu\nu\beta_j}(\mathbf{q,k_j-q,k_j})\frac{k_BT}{m_{\beta_j}}~.
\label{eq:kn}
\end{align}
Here $\mathcal{V}_{\alpha\beta\gamma}(\mathbf{q,k-q,k})$ is the static vertex function, which remains equal to the standard one of MCT~\cite{Voigtmann2003} and physically represents the coupling strength among different wavevectors. Explicitly, the vertex function reads
\begin{equation}
   \mathcal{V}_{\alpha\beta\gamma}(\mathbf{q,k-q,k})= \delta_{\beta\gamma}\bq\cdot\bk c_{\alpha\gamma}(q)+\delta_{\alpha\gamma}(\bk-\bq)\cdot\bk c_{\beta\gamma}(|\bk-\bq|),
\end{equation}
with the direct correlation function $c_{\alpha\beta}(q)$. It relates to the static structure factors via the Ornstein-Zernike equation $c_{\alpha\beta}(q)=\rho^{-1}(\delta_{\alpha\beta}/x_{\alpha}-(\bS^{-1}(q))_{\alpha\beta}$ where $\rho=N/V$ is the number density of the system \cite{Hansen2006}. Finally, we note that in this derivation we have neglected the so-called projected dynamics and we have ignored the off-diagonal correlations, similar to what is done in conventional MCT and single-component GMCT~\cite{Reichman2005,Gotze1992,Liesbeth2018front}.  
Equation \ref{eq:order-n} is subject to the initial boundary conditions $\dot{F}^{(n)}_{\{\alpha_i\};\{\beta_i\}}(\{k_i\},t=0)=0$ and 
$F^{(n)}_{\{\alpha_i\};\{\beta_i\}}(\{k_i\},t=0)=S^{(n)}_{\{\alpha_i\};\{\beta_i\}}(\{k_i\})$ (Eq.\ \ref{eq:Sn}) for all $\{\alpha_i\}$,$\{\beta_i\}$, and $\{k_i\}$.

In principle the above hierarchical equations can be solved up to arbitrary order $n$, but in practice we must apply a suitable closure at finite order $n_\mathrm{max}$ to obtain numerically tractable results. We use the following mean-field closure at level $n_\mathrm{max}>2$~\cite{Janssen2015a}
\begin{eqnarray}
K^{(n_\mathrm{max}-1)}_{\{\alpha_i\};\{\beta_i\}}(\{k_i\},t)
\approx
\frac{1}{n_\mathrm{max}-2}\sum_j
K^{(n_\mathrm{max}-2)}_{\{\alpha_i\}/\alpha_j;\{\beta_i\}/\beta_j}(\{k_i\}/k_j,t)
 F^{(1)}_{\alpha_j;\beta_j}(k_j,t) \; ~.
\label{eq:closure}
\end{eqnarray}
Note that this makes use of the permutation invariance of all wavenumber arguments $\{k_1,\hdots,k_n\}$. 
In terms of the intermediate scattering function the closure is equivalent to the factorization approximation $F^{(n_\mathrm{max})}(t)\sim F^{(n_\mathrm{max}-1)}(t)\times F^{(1)}(t)$.
Hence at $n_{max}=2$ we obtain the same closure as in standard multi-component MCT~\cite{Voigtmann2003}.
As a reminder, most of the problems with standard MCT come from the fact that such a factorization closure is too strong and unjustified~\cite{gotze2008complex,Gotze1992,Reichman2005}.
Multi-component GMCT allows us instead to shift the closure to a larger $n_{max}>2$, meaning that the correlations $F^{(n')}(t)$ of order $n'<n_{max}$ are not factorized and are more correctly described.
This approach has already been shown to be beneficial in single-component glassy systems~\cite{SzamelPRL2003,Wu2005,Janssen2015a,Luo2020_1,Luo2020_2}, and in this paper we demonstrate that a similar improvement can be gained for binary systems.

\section{Numerical details}
\subsection{Numerical solution of GMCT}
Since the system is isotropic and invariant under rotations, we use bipolar coordinates to transform the three-dimensional integrals over $\bq$ that appear in any memory function of the hierarchy (Eq.~\ref{eq:kn}) as a double integral over $q = |\bq|$ and $p = |\bk - \bq|$. 
Then, $q$ is discretized over a uniformly spaced grid of $N_k=70$ points $q = q_0 + \hat{q} \Delta q$ with $\hat{q} = 0, 1, \ldots, N_k-1$ and $\Delta q=40/N_k$. In the Supplementary Information we show that a grid of $N_k=70$ wavenumbers is sufficiently converged to predict the MCT critical point, at least for binary Percus-Yevick hard spheres. This choice of parameters allows us to replace the double integral by Riemann sums
\begin{equation}
  \int_0^\infty dk  \int_{|q-k|}^{q+k} dp
  \; \rightarrow \;
  (\Delta q)^2 \sum_{\hat{k}=0}^{N_k-1} \sum_{\hat{p} = |\hat{q}-\hat{k}|}^{\min[N_k-1,\hat{q}+\hat{k}]} .
  \label{eq:discreteqintegral}
\end{equation}
Following Ref.~\cite{Franosch1997} we set $q_0=\Delta q/2$ in order to prevent any possible divergence for $q\xrightarrow[]{}0$.

To obtain the time-dependent solutions of Eq.~\ref{eq:order-n}, we start with a Taylor expansion around $t=0$ for all dynamical correlation functions $F^{(n)}$ up to the level $n_\mathrm{max}-1$; the correlator at the highest level, $F^{(n_\mathrm{max})}$, 
follows from our closure relation, Eq.~\ref{eq:closure}.  We then integrate Eq.~\ref{eq:order-n} in time using Fuchs' algorithm~\cite{Fuchs1991}, where the first $N_t=64$ time points are calculated with a step size of $\Delta t = 10^{-6}$, and $\Delta t$ is subsequently doubled every $N_t/2$ points. At each point in time, we iteratively update the wavevector-dependent memory kernels (Eq.~\ref{eq:kn}) for all $n\leq n_\mathrm{max}$ until convergence. Note that in these GMCT equations, the partial static structure factors $S^{\alpha\beta}(k)$ enter both in the initial boundary conditions for  $F^{(n)}$, as well as in the static vertices and the matrices $J^{(n)}$.
In summary, at any given $(T,\rho)$ we only require $S^{\alpha\beta}(k)$ as input to predict the microscopic relaxation dynamics of the system. While our GMCT framework gives access to all multi-point dynamical density correlations up to order $n_\mathrm{max}$, in the following we shall restrict the discussion to the intermediate scattering function $F^{\alpha\beta}(k,t)=F^{(1)}_{\alpha;\beta}(k,t)$.

\subsection{Numerical simulations}
We use multi-component GMCT to predict the glassy behavior of two binary mixtures: the Kob-Andersen binary Lennard-Jones (LJ) mixture~\cite{Kob1994} and its Weeks-Chandler-Anderson truncation (WCA) ~\cite{wca1971}.
Both are three-dimensional $80:20$ mixtures of particles $A:B$ which interact with each other via the following potential
\begin{equation}
    V_{\alpha\beta}(r)=
    \begin{cases}
    4\epsilon_{\alpha\beta}\left[\left(\frac{\sigma_{\alpha\beta}}{r}\right)^{12}-\left(\frac{\sigma_{\alpha\beta}}{r}\right)^{6} +C_{\alpha\beta}\right]~,\qquad r\le r_{\alpha\beta}^c~,
    \\
    \qquad\qquad\qquad 0~,\qquad\qquad\qquad\qquad\ \ r>r_{\alpha\beta}^c~.
    \end{cases}
\end{equation}
Here the cutoff radius $r^c_{\alpha\beta}$ is $2.5\sigma_{\alpha\beta}$ for LJ, while it corresponds to the potential minimum for WCA~\cite{wca1971}. The constant $C_{\alpha\beta}$ ensures that $V_{\alpha\beta}(r^c_{\alpha\beta})=0$.
We use $\epsilon_{AA}=1, \epsilon_{AB}=1.5, \epsilon_{BB}=0.5, \sigma_{AA}=1, \sigma_{AB}=0.8, \sigma_{BB}=0.88$ to obtain good glass-forming mixtures~\cite{Kob1994}.

In order to calculate the relevant quantities we need for a comparison with multi-component GMCT, we perform molecular dynamics simulations in the NVE ensemble using HOOMD-blue~\cite{anderson2008general}. We properly equilibrate both systems at different densities $\rho$ and temperatures $T$, and use $N=1000$ particles. The density is tuned via the size of the periodic box $L$, and all parameters and results are reported in terms of reduced WCA units~\cite{wca1971}. 
From the simulation trajectories, we calculate the partial static structure factors $S^{\alpha\beta}(k)$ and the collective intermediate scattering functions $F^{\alpha\beta}(k,t)$.
For the multi-component GMCT calculations, we use the simulated $S^{\alpha\beta}(k)$ as the input of Eq.~\ref{eq:order-n} to predict the theoretical $F^{\alpha\beta}(k,t)$. 
In the next section we compare the output of multi-component GMCT with the $F^{\alpha\beta}(k,t)$ obtained from simulation, and show that multi-component GMCT becomes progressively closer to the simulated glass transition temperature as we increase the level of the GMCT hierarchy.

\section{Results and discussion}

\subsection{From structure to dynamics}
\begin{figure}
\centering
  \begin{minipage}[b]{0.49\textwidth}
    \includegraphics[width=\textwidth]{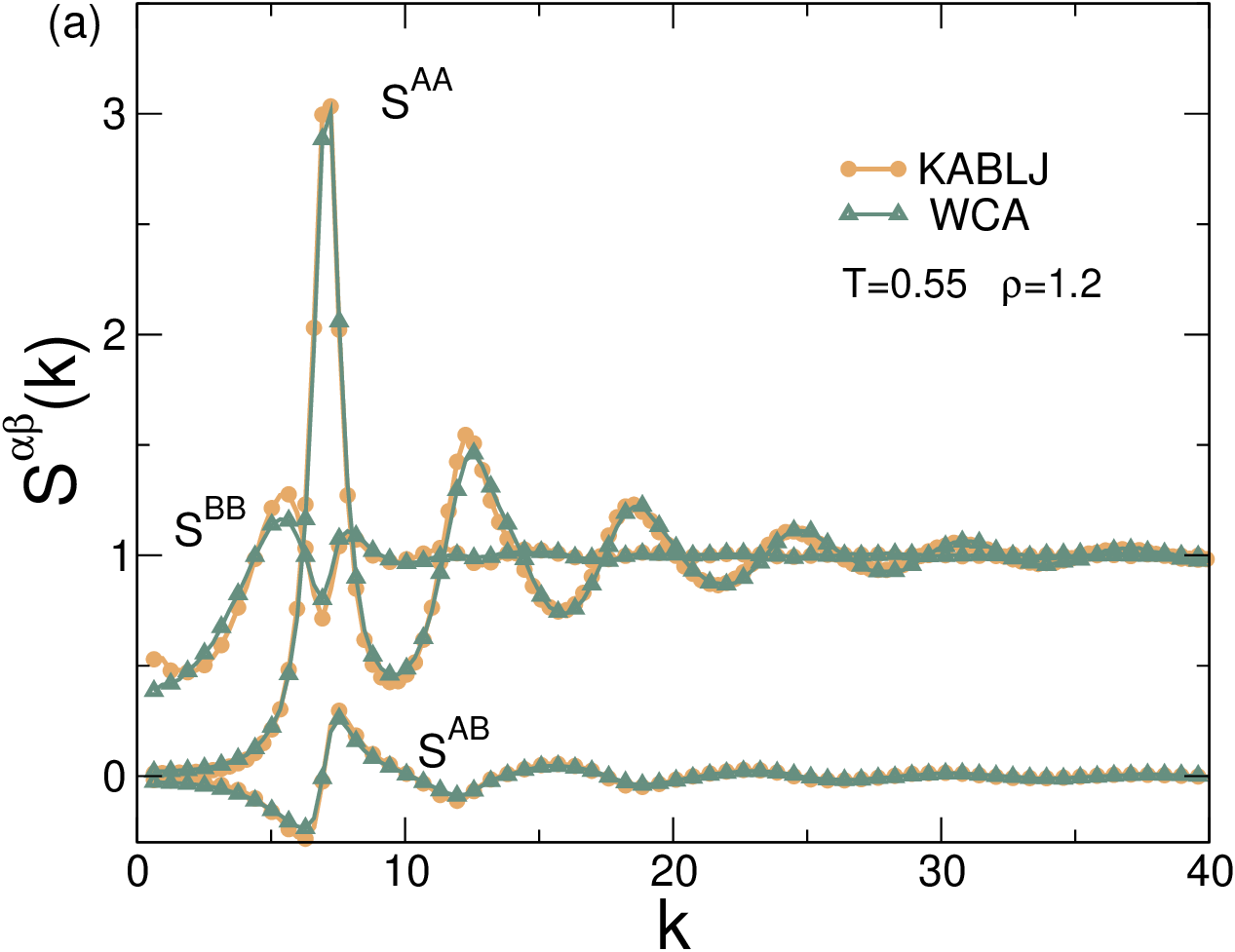}
  \end{minipage}
  \hfill
  \begin{minipage}[b]{0.49\textwidth}
    \includegraphics[width=\textwidth]{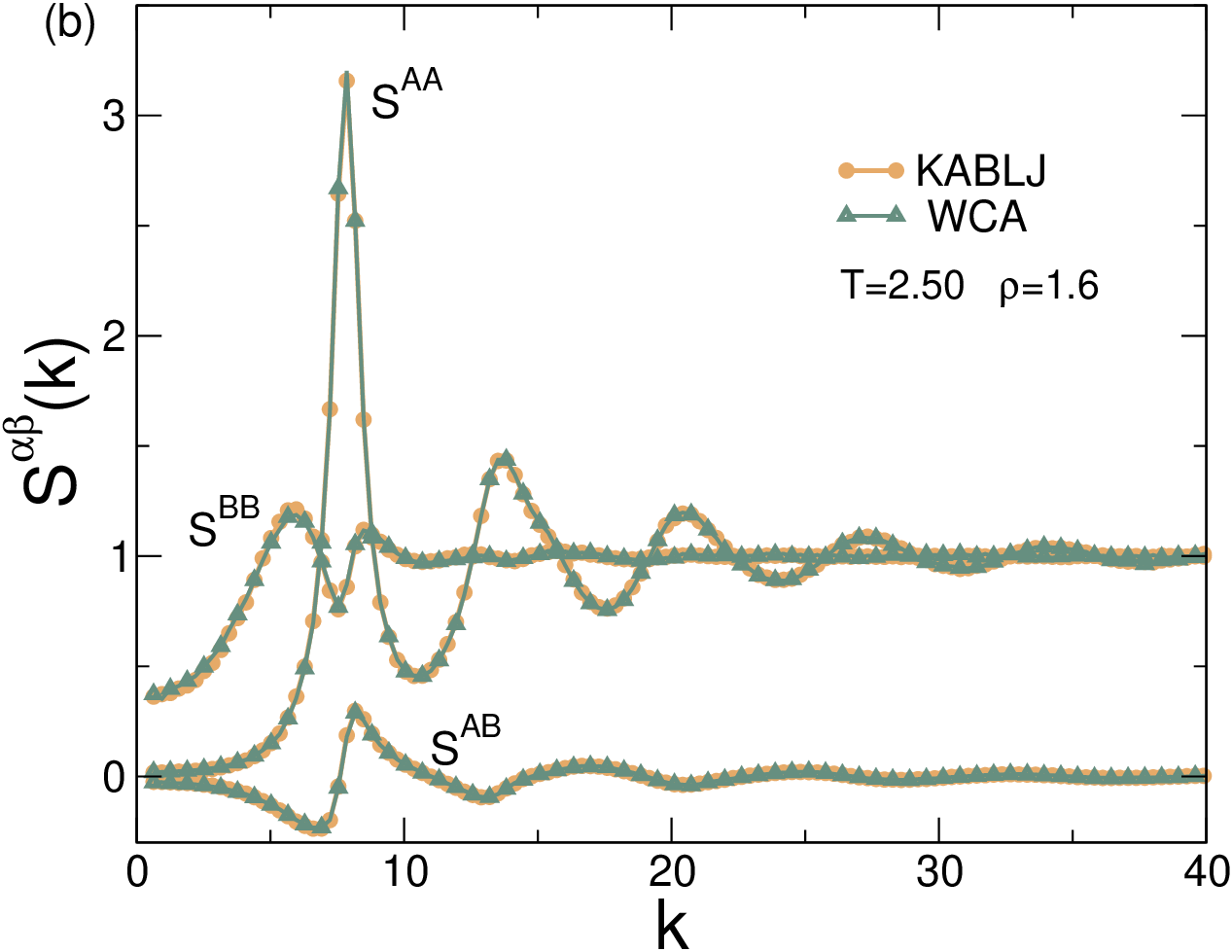}
  \end{minipage}
  \begin{minipage}[b]{0.49\textwidth}
    \includegraphics[width=\textwidth]{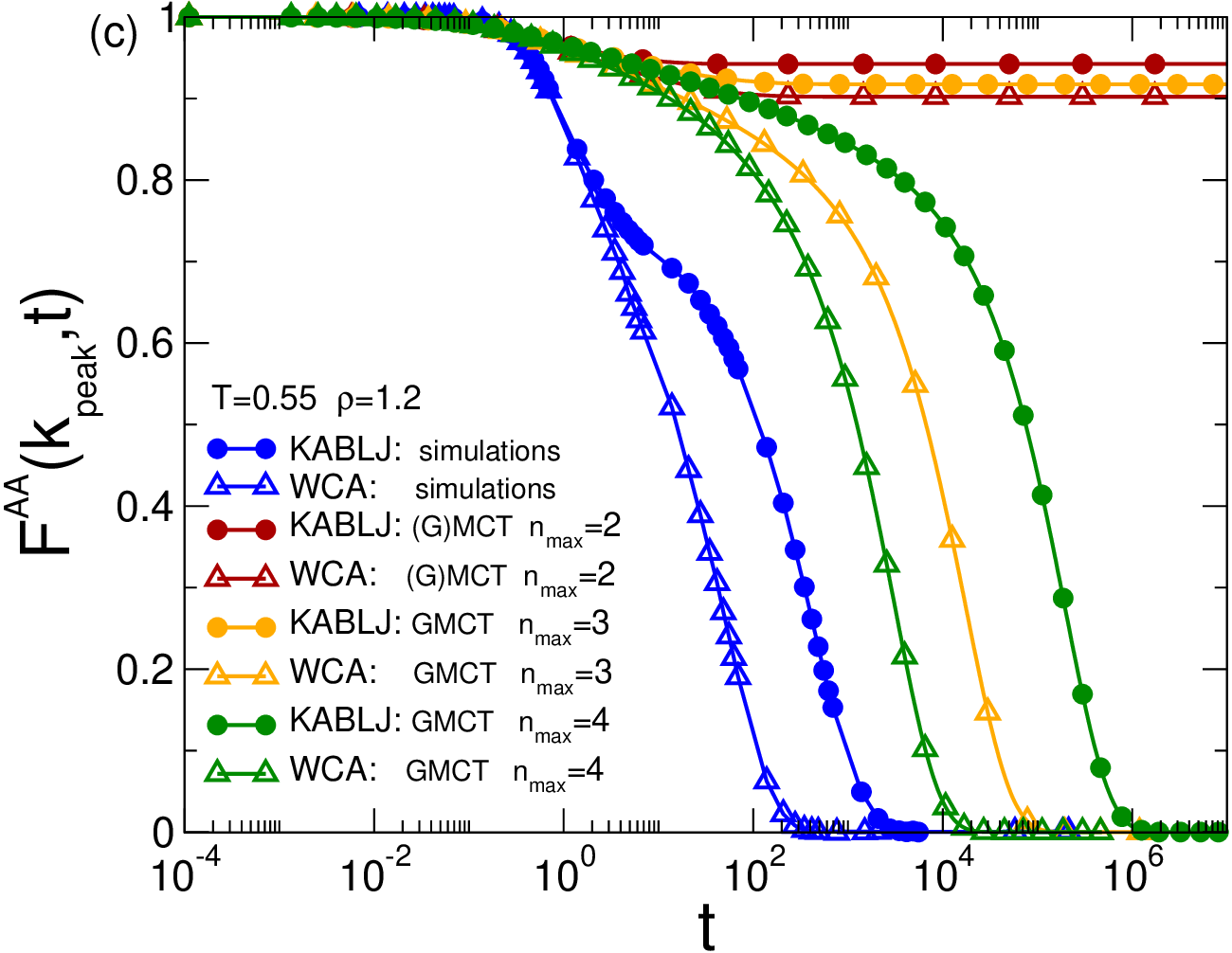}
  \end{minipage}
  \hfill
  \begin{minipage}[b]{0.49\textwidth}
    \includegraphics[width=\textwidth]{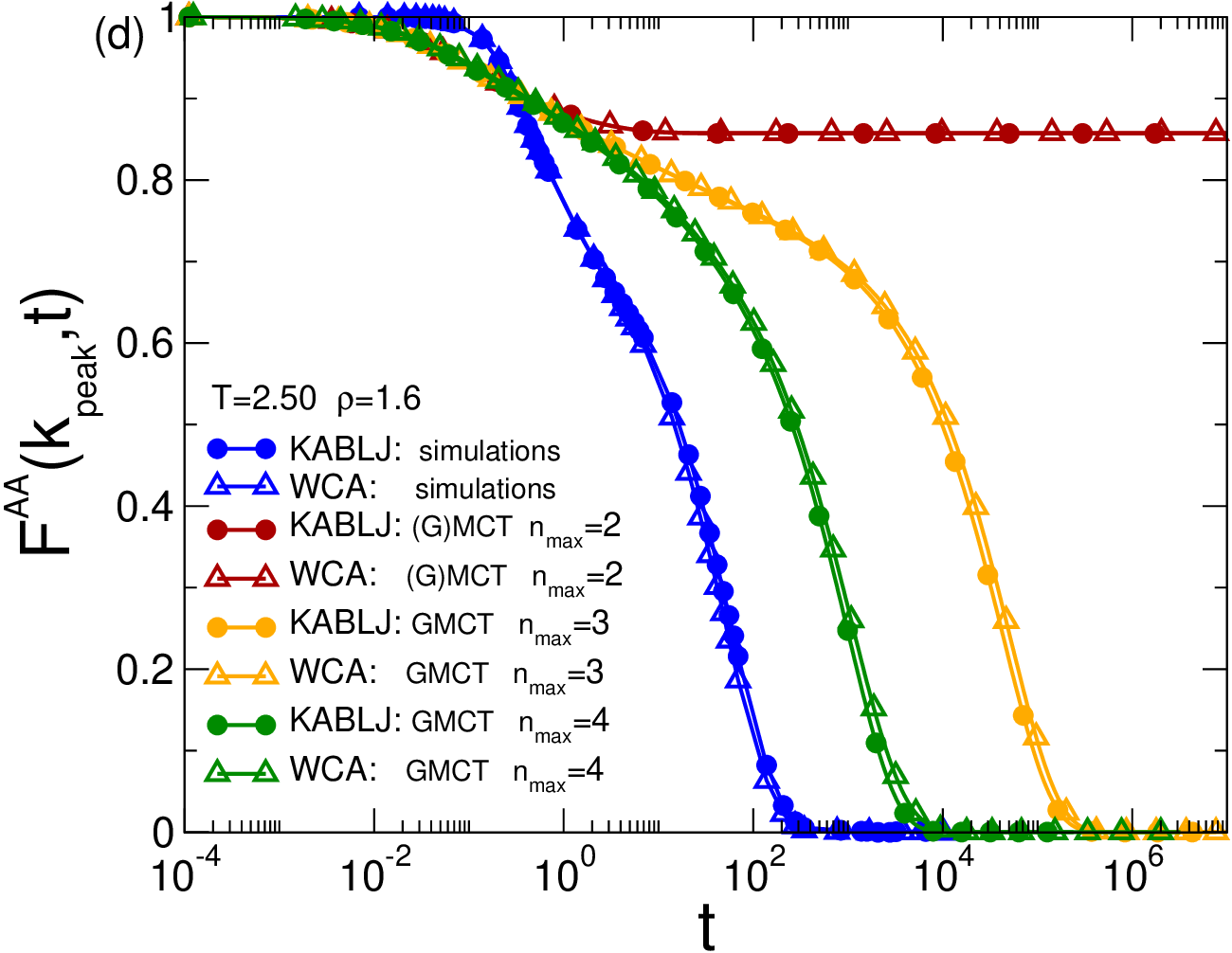}
  \end{minipage}
    \centering
    \caption{Structure and dynamics of supercooled binary LJ and WCA mixtures at $T=0.55, \rho=1.2$ and $T=2.50, \rho=1.6$. 
    In panels (a) and (b) we compare the partial static structure factors of binary LJ (yellow) and binary WCA (gray). In (c) and (d) we show the dynamics in the supercooled regime of the component $F^{AA}(k,t)$, which is the term that dominates the dynamics since $80\%$ of the system is type $A$. The wavenumber $k=k_\mathrm{peak}$ corresponds to the maximum of $S^{AA}(k)$. The different curves in panels (c) and (d) show $F^{AA}(k,t)$ measured from MD simulations (blue), binary MCT (red), binary GMCT with $n_\mathrm{max}=3$ (orange), and binary GMCT with  $n_\mathrm{max}=4$ (green). When increasing the level of the GMCT hierarchy $n_\mathrm{max}$, the $F^{AA}(k,t)$ predicted by multi-component GMCT tends to converge to the simulation results.   
    }
    \label{fig:SF_mgmct}
\end{figure}
The strength of GMCT is its capability of predicting dynamics from statics. The first result that we show underlines the sensitivity of GMCT to small variations in the static structure.
In Fig.~\ref{fig:SF_mgmct}(a) we compare the partial structure factors of the binary LJ (yellow) and binary WCA (gray) systems at density $\rho=1.2$ and temperature $T=0.55$, which corresponds to low density in the supercooled regime. Notice that all the components of the structure factor are very similar between the two mixtures, consistent with previous simulations~\cite{Coslovich2012,Berthier2011a,Berthier2010,Landes2020}. 
However, as shown in Fig.~\ref{fig:SF_mgmct}(c), and also in agreement with earlier studies~\cite{Coslovich2012,Berthier2011a,Berthier2010,Landes2020}, the simulated relaxation dynamics of the two systems (blue curves) differ significantly. In particular, the structural relaxation of $F^{AA}(k_\mathrm{peak},t)$, with $k=k_\mathrm{peak}$ corresponding to the main peak of $S^{AA}(k)$, is approximately one order of magnitude slower for the LJ mixture. This disparity in dynamics also becomes more pronounced when decreasing $T$.

Importantly, binary MCT can only partly account for these dynamical differences based on the input static structure factors, and furthermore the standard theory cannot reach quantitative accuracy for either system at any given temperature~\cite{Nagele1999a,Voigtmann2003,Gotze2003,Weysser2010}. Indeed, it is also demonstrated in Fig.~\ref{fig:SF_mgmct}(c) that binary MCT (red curves) fails to predict the correct dynamics at this temperature and density, erroneously predicting a non-ergodic glass phase for both systems. Our multi-component GMCT framework, on the other hand, better approaches the simulated long-time dynamics \textit{from the same $S^{\alpha\beta}(k)$ as input} as we increase the level of the hierarchy $n_\mathrm{max}$. In particular, note that the highest considered GMCT closure level, $n_\mathrm{max}=4$ (green curves), correctly yields an ergodic phase for both systems, with the LJ mixture having one to two orders of magnitude slower relaxation dynamics than the WCA mixture. This prediction is in good qualitative agreement with simulation. 


When the density is high ($\rho=1.6$) the attraction that distinguishes LJ from WCA is less significant, since all particles predominantly probe only the short-range repulsive regime. In fact we see in Fig.~\ref{fig:SF_mgmct}(b) that all the components of $S^{\alpha\beta}(k)$ are virtually identical among the two mixtures.
The dynamics in this supercooled regime, reported in Fig.~\ref{fig:SF_mgmct}(d), is also almost the same for the simulated mixtures. Notice that the value of $T=2.5$ corresponds to approximately $1.5T_g$, and it is comparable to the value of $T=0.55$ at $\rho=1.2$ of Fig.~\ref{fig:SF_mgmct}(c). Similarly, every level of the binary GMCT hierarchy also predicts almost indistinguishable dynamics at this density.
However, it is once again noticeable that a higher $n_\mathrm{max}$ makes multi-component GMCT converge towards the simulations.

Overall, Fig.~\ref{fig:SF_mgmct} clearly shows that small differences in the structure are captured by multi-component GMCT and amplified to predict the dynamics in the glassy regime. While on the one hand this sensitivity of the theory means that a high precision is required when measuring the input-$S(k)$, on the other hand this supports the idea that important information about the dynamics is already enclosed in static $2-$point density correlations~\cite{Coslovich2012,Landes2020}.

\subsection{The role of polydispersity}
\begin{figure}
    \centering
    \includegraphics[width=0.75\textwidth]{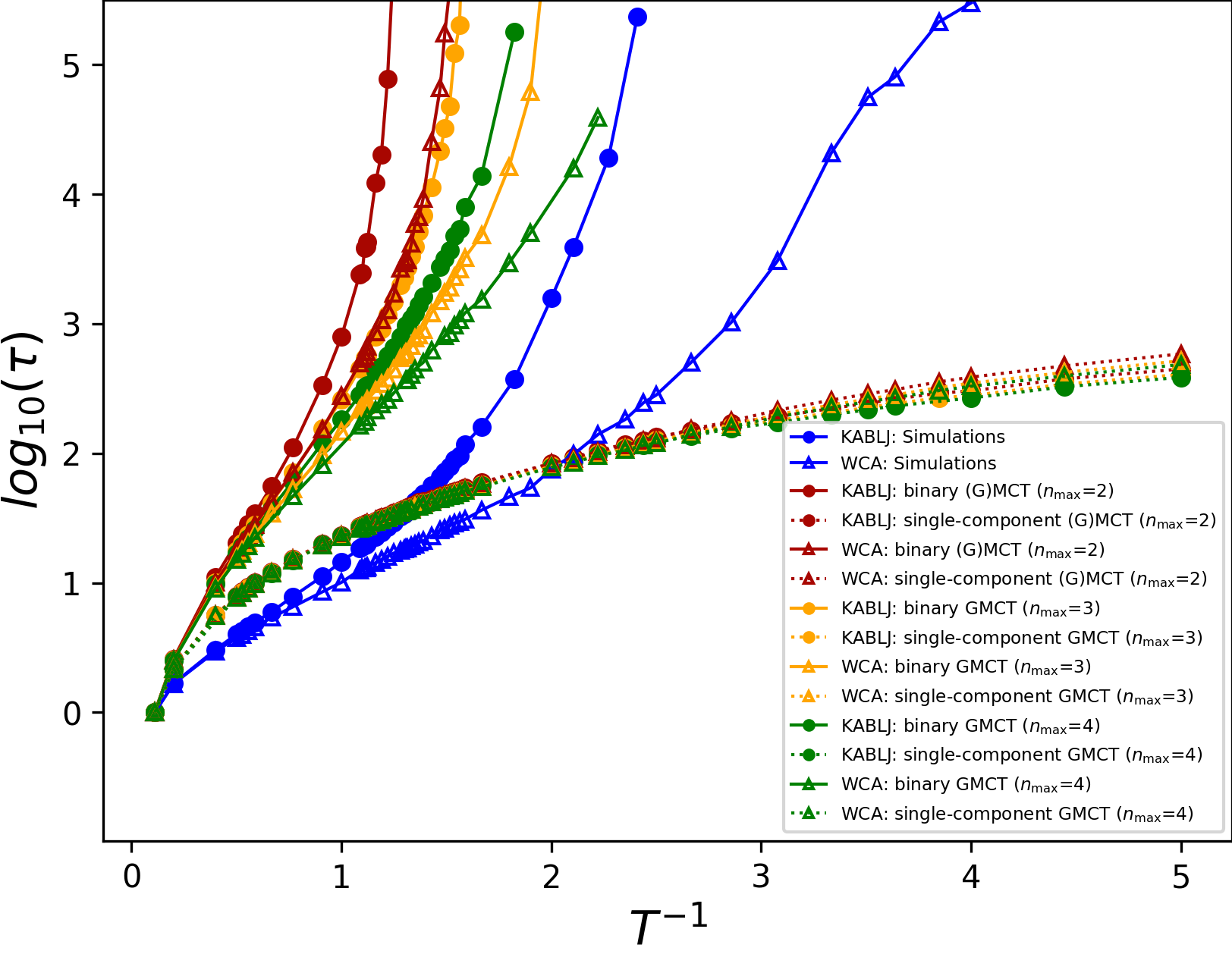}
    \caption{Relaxation time as a function of inverse temperature from simulations (blue), single-component (dashed lines)
    and multi-component (continuous line) (G)MCT, for binary LJ and WCA at $\rho=1.2$.
    The data show that single-component GMCT does capture only a very weak slowdown upon supercoooling and does not show any trace of a critical point, nor any significant improvement for larger $n_\mathrm{max}$. The results of multi-component GMCT are discussed more in detail in the next figure; here we only show that multi-component GMCT reproduces the binary simulations more realistically than single-component GMCT. 
    }
    \label{fig:singcomp}
\end{figure}
By extending the framework of GMCT to include multiple components, we can directly account for polydispersity, i.e.\ the heterogeneity of sizes of molecules or particles in a mixture. It is ubiquitous in experiments at the colloidal scale because two particles are hardly equal and in the context of glasses it is 
also useful to avoid crystallization~\cite{Angell1995}. Furthermore, it has been shown that even in simulations where monodispersity is possible, it can be beneficial  to use polydispersity in order to employ algorithms such as Monte Carlo swaps which can significantly improve the performance of computations~\cite{Ninarello2017a}.

If the degree of polydispersity is small it has been shown that single-component GMCT is capable of very accurate predictions~\cite{Janssen2015a}. However, for highly polydisperse systems or complex architectures~\cite{Ciarella2019pnas,Frey2015,chong2007structural,Baschnagel_2005} single-component theories require a pre-averaging of the structure. This can severely influence their predictions.
In particular, since (G)MCT is very sensitive to the value of the main peak of the static structure factor~\cite{Ciarella2019pnas}, averaging the $AA$ correlation with the $AB$ and $BB$ components inevitably leads to a decrease of such peak which, in turn, alters the results of (G)MCT. 

In Fig.~\ref{fig:singcomp} we examine the consequences of approximating a highly polydisperse system, i.e.\ our binary LJ and WCA mixtures, as being effectively monodisperse.
We report the relaxation time $\tau$ as a function of the inverse temperature at $\rho=1.2$, comparing simulations to single-component and multi-component GMCT; for single-component GMCT we use the average  static structure factors $S(k)$ as input, whereas for multi-component GMCT we explicitly distinguish between all the partial components $S^{\alpha\beta}(k)$.
The relaxation time $\tau$ 
is defined as
\begin{equation}
    F^{AA}(k_\mathrm{peak},t=\tau)=\frac{F^{AA}(k_\mathrm{peak},0)}{e}~,
    \label{eq:tau}
\end{equation}
which grows rapidly during supercooling towards the glass transition temperature $T_g$~\cite{Angell1995}.
We find that single-component (G)MCT significantly underestimates the critical glass-transition temperature for both systems. In particular, at the supercooled temperatures where the simulated $\tau$ reaches a value of $\tau\sim10^4-10^5$ (i.e.\ near the simulated $T_g$), we find that our single-component (G)MCT approximation yields a relaxation time that is almost three orders of magnitude too low. 
This underestimation of the glassy dynamics is also consistent with MCT studies of polymeric systems that use pre-averaged static structure factors~\cite{Ciarella2019pnas,Frey2015,chong2007structural,Baschnagel_2005}. 
Moreover, note that the qualitative shape of the $\tau(T)$ curves predicted by single-component theory also deviates markedly from the simulation results, and that little improvement is gained by increasing $n_\mathrm{max}$. 

By contrast, when properly taking into account the binary nature of both systems, multi-component (G)MCT yields predictions that more closely resemble the $\tau(T)$ simulation curves, at least on a qualitative level. We also see that the multi-component theory in fact overestimates the critical temperature, with the highest overestimation found for the lowest $n_\mathrm{max}$. This general tendency to overestimate the glassiness is also consistent with other
multi-component~\cite{Voigtmann2003,Weysser2010} and standard MCT~\cite{Reichman2005} calculations.
Overall, these results underline the fact that non-trivial couplings exist in the structure and dynamics of multi-component glassy mixtures, highlighting the need to explicitly account for polydispersity in such systems. 

\subsection{Relaxation time}
\begin{figure}
\centering
  \begin{minipage}[b]{0.49\textwidth}
    \includegraphics[width=\textwidth]{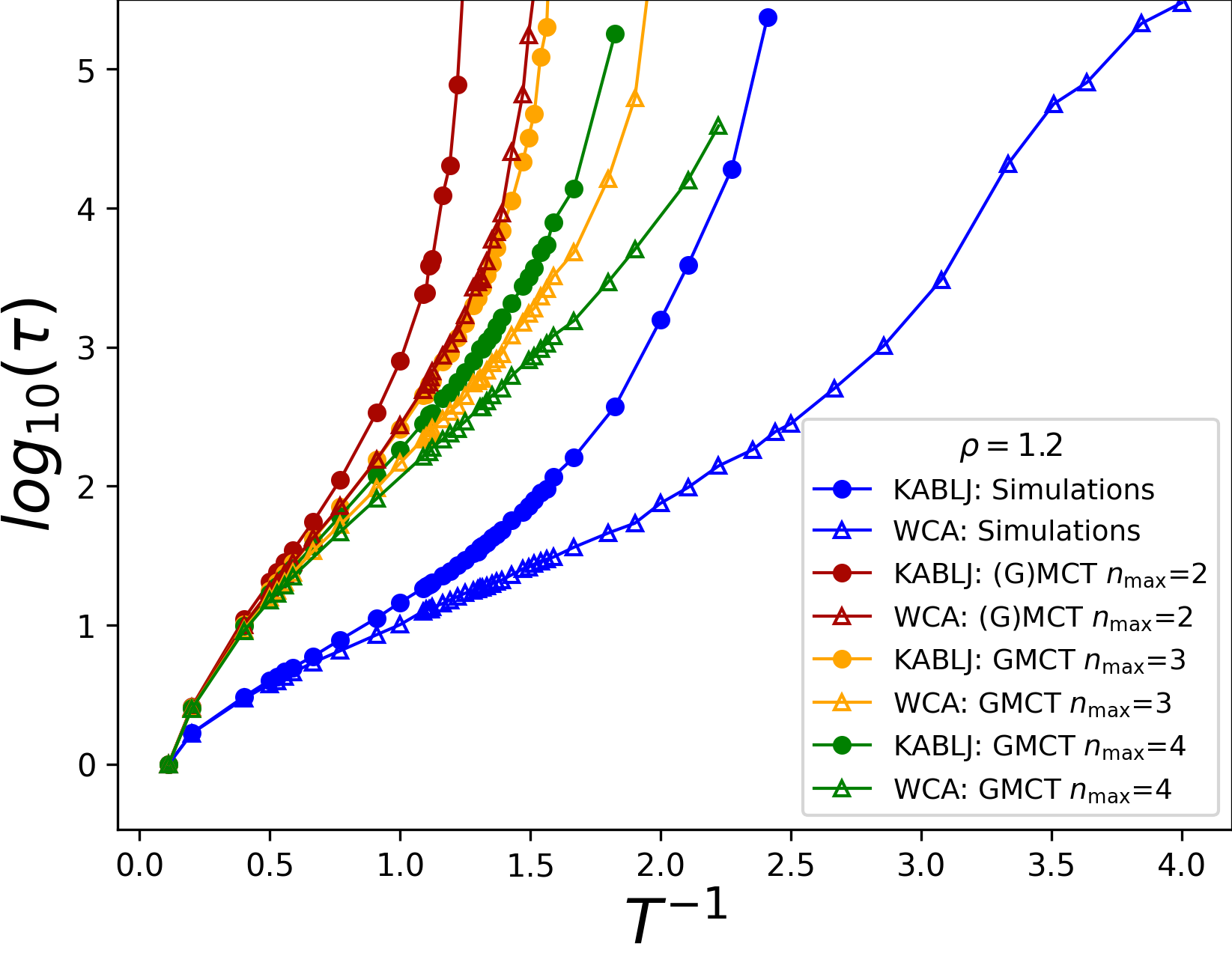}
  \end{minipage}
  \hfill
  \begin{minipage}[b]{0.49\textwidth}
    \includegraphics[width=\textwidth]{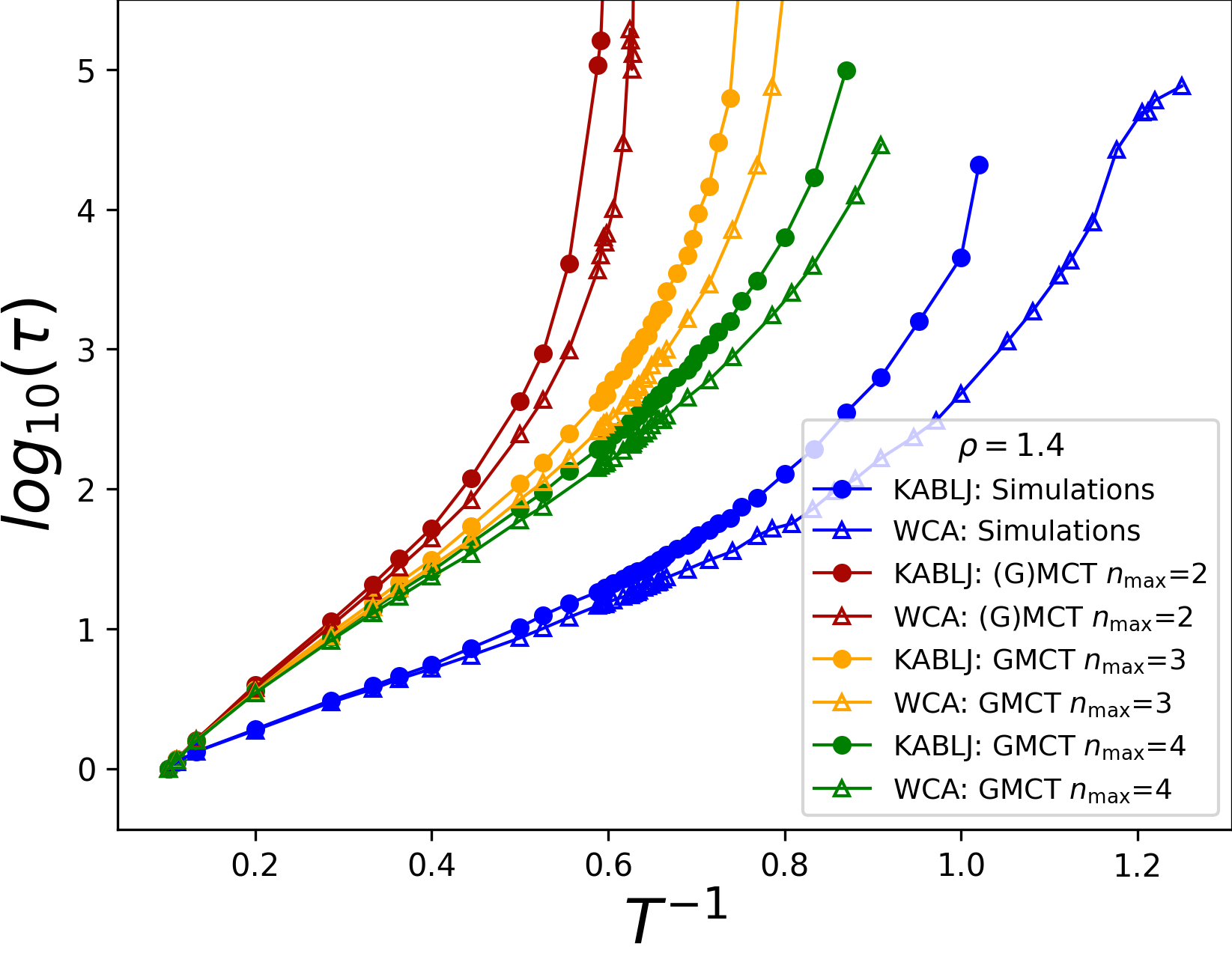}
  \end{minipage}
    \includegraphics[width=0.49\textwidth]{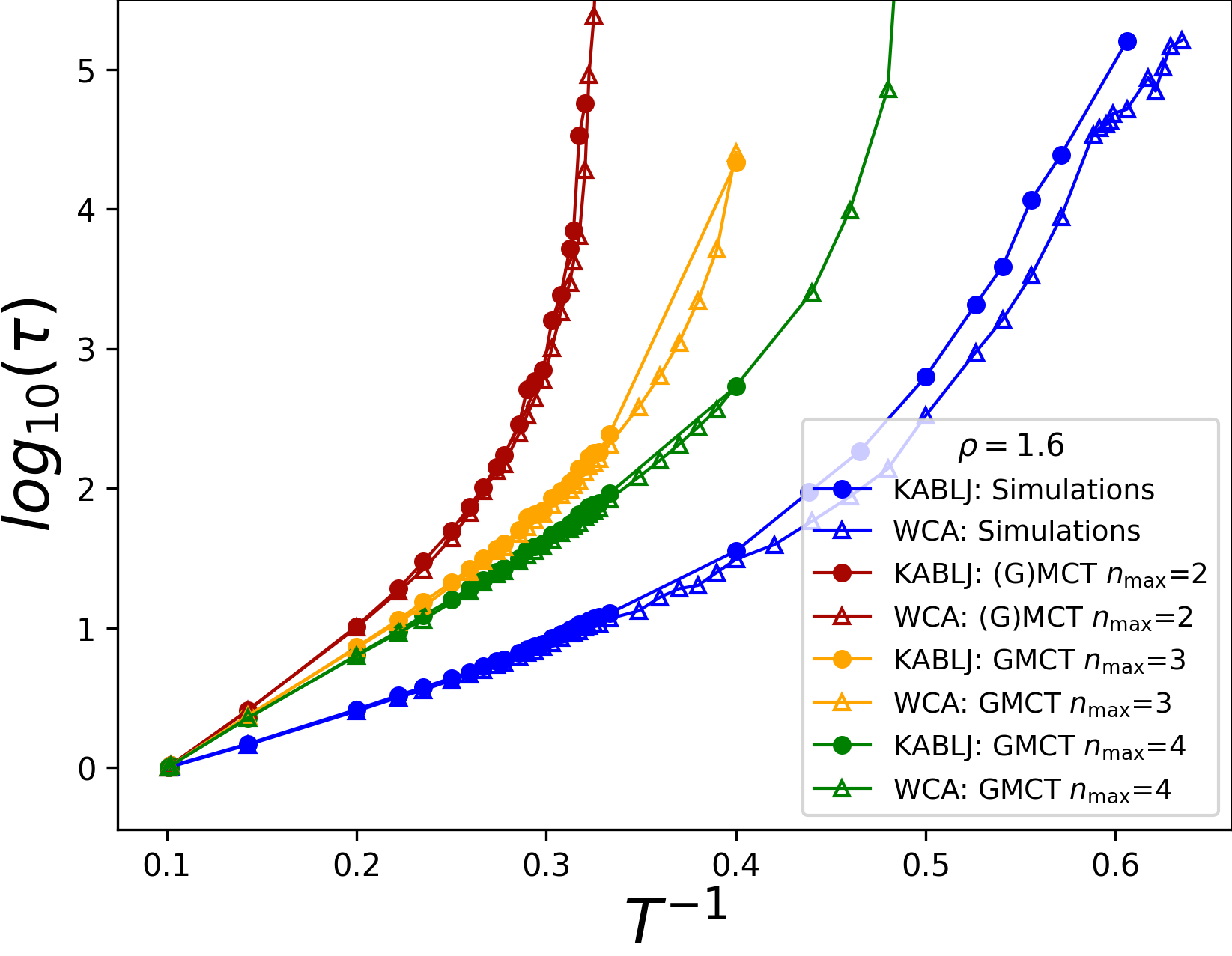}
    \centering
    \caption{Relaxation time as a function of inverse temperature from simulations (blue) and multi-component (G)MCT, for binary LJ and WCA mixtures at $\rho=1.2, 1.4$, and $1.6$.
    The relaxation time is evaluated from $F^{AA}(k_\mathrm{peak},t)$, corresponding to the majority species $A$ and the  wavenumber $k=k_\mathrm{peak}$ where $S^{AA}(k)$ has its maximum. 
    The data show that by increasing the GMCT closure level, the predictions of multi-component GMCT become increasingly more accurate. }
    \label{fig:mgmgt}
\end{figure}
We proceed by comparing our numerical simulations with the predictions of multi-component GMCT.
The comparison is summarized in Fig.~\ref{fig:mgmgt} where we report the relaxation time $\tau$ as a function of the inverse temperature for three different bulk densities.
It should be noted that here we  solely focus on $\alpha=\beta=A$ when determining $\tau$, because particles of type $A$ constitute $80\%$ of the system and therefore dominate the dynamics. 
Furthermore, as before, we set $k=k_\mathrm{peak}$, corresponding to the maximum of $S^{AA}(k)$, thus focusing on the slowest modes in the system~\cite{Reichman2005,Kob2002,Liesbeth2018front}.

The results in Fig.~\ref{fig:mgmgt} show that MCT (red curves, corresponding to GMCT with closure level $n_\mathrm{max}=2$) overestimates the value of $T_g$ obtained from simulations (blue) as expected~\cite{Berthier2010,Berthier2011a}. 
However, if we increase the level of the hierarchy to $n_\mathrm{max}=3$ (orange) and then $n_\mathrm{max}=4$ (green), the accuracy increases and the critical point of GMCT manifestly converges towards the simulations. This uniform convergence of the multi-component theory with increasing $n_\mathrm{max}$ is also  consistent with earlier findings from single-component GMCT~\cite{Janssen2015a,Janssen2016a,Luo2020_1,Luo2020_2}.

From the data in Fig.~\ref{fig:mgmgt}, it is particularly noteworthy that at $\rho=1.2$, where the difference between the simulated LJ and WCA dynamics is the largest, higher-order multi-component GMCT  becomes increasingly better at distinguishing between the two mixtures~\cite{Berthier2009,Berthier2011a}. 
Here it is important to recall that for each temperature and density considered, all our GMCT calculations use the same $S^{\alpha\beta}(k)$ as input, regardless of the chosen $n_\mathrm{max}$. The fact that increasing $n_\mathrm{max}$ leads to better dynamical predictions, and perhaps might even become (near-)exact in the limit of $n_\mathrm{max}\xrightarrow[]{}\infty$~\cite{Janssen2016a}, clearly suggests that static 2-point correlations already constitute an important indicator of glassiness--provided that the appropriate dynamical framework is used to translate structure into dynamics. The importance of structural pair correlations  has also been verified recently through agnostic machine learning methods~\cite{Landes2020,Schoenholz2016,Cubuk2015,paret2020}, and with the here presented work we can now place this result on a firmer, first-principles-based theoretical footing. 


\subsection{Role of attraction}
\begin{figure}
    \centering
    \includegraphics[width=0.49\columnwidth]{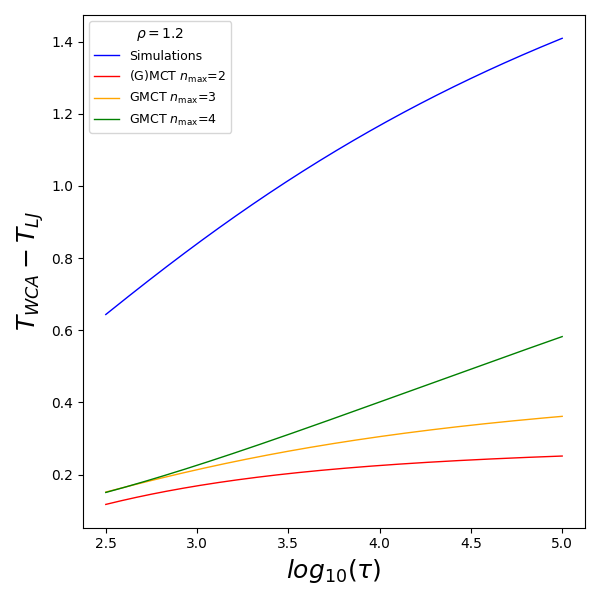}
    \includegraphics[width=0.49\columnwidth]{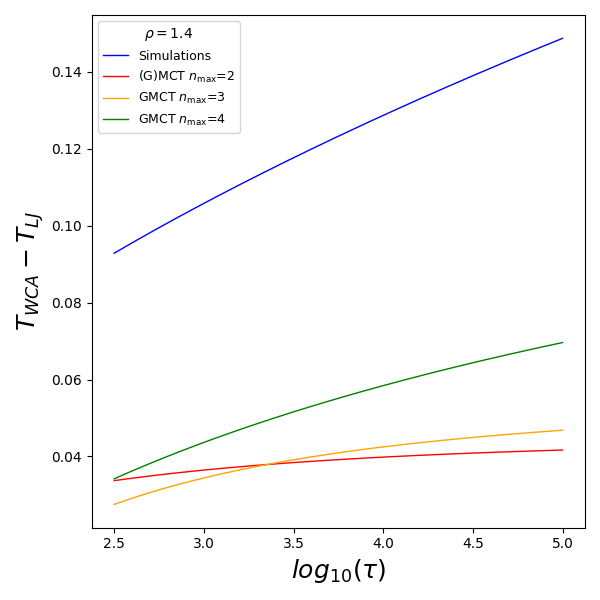}
    \caption{Effect of the attraction in binary LJ mixtures relative to repulsive WCA, measured in terms of the temperature difference $T_{WCA}-T_{LJ}$ at which both systems yield the same relaxation time $\tau$. The higher the closure level of binary GMCT, the larger this temperature difference becomes. Hence we can conclude that higher order GMCT better recognizes the role of attraction via the corresponding changes in the static structure factor.}
    \label{fig:attraction}
\end{figure}
The static structure factor $S(k)$ has also been shown to contain information about higher order static correlations~\cite{Coslovich2012,Zhang2020}. 
However, if we use the relevant $S^{\alpha\beta}(k)$ as the main input of standard MCT, the theory is not able to efficiently distinguish between LJ and WCA mixtures~\cite{Berthier2010,Berthier2011a} (also see Fig.~\ref{fig:mgmgt}). This implies that at least on the MCT level, the role of attractive particle interactions in supercooled liquids is not adequately captured.

We show here that higher order (multi-component) GMCT is more sensitive to small differences in $S(k)$ and thus the theory is able to recursively recognize better the role of attraction.
To support this claim, we compare the LJ and WCA dynamics at different temperatures $T_{LJ}$ and $T_{WCA}$, respectively, where the temperatures are defined such that they yield the \textit{same} relaxation time $\tau$. 
In Fig.~\ref{fig:attraction} we report the measured temperature difference $T_{WCA}-T_{LJ}$ as a function of the relaxation time $\tau$.
This analysis is based on a power law fitting $\tau \sim (T-T_0)^{-\gamma} + A_0$, where the  parameters $A_0,T_0$, and $\gamma$ are fitted to best approximate Fig.~\ref{fig:mgmgt} for each value of $\rho$. All the numerical values are reported in the Supplementary Information (Table I).
It can be seen in Fig.~\ref{fig:attraction} that the temperature difference extracted from the simulations (blue) becomes progressively larger as $\tau$ increases, indicative of the markedly different supercooled LJ and WCA dynamics. 
In standard MCT this difference is not properly captured; in fact, binary MCT (red) predicts that the temperature difference is always small and almost constant.
However, when we increase the closure level of binary GMCT to $n_\mathrm{max}=3$ (yellow) and  $n_\mathrm{max}=4$ (green) the difference $T_{WCA}-T_{LJ}$ becomes larger and, similarly to the simulations, it grows approaching the glass transition.
We therefore conclude that higher order GMCT can extract more information from $S(k)$ and hence it is able to better recognize the role of attraction in the emergent supercooled dynamics.

\subsection{Conclusions}
In this paper we have derived generalized MCT for multi-component systems, thus extending the earlier version of the theory~\cite{SzamelPRL2003,Wu2005,Janssen2015a,Luo2020_1,Luo2020_2} to the case of mixtures with an arbitrary number of species. 
The theory seeks to predict the microscopic relaxation dynamics of glassy mixtures in a fit-parameter-free manner using the static structure factors as its main input.
Its hierarchical structure of nested integro-differential equations can be closed and solved self-consistently at any order $n$. 
The predictive power of the theory manifestly increases for larger $n$, providing a promising, and systematically improvable framework to ultimately achieve an accurate description of the elusive structure-dynamics link in glass-forming liquids. 

We have used the newly derived multi-component GMCT to describe the glassy dynamics of three-dimensional Kob-Andersen LJ and WCA binary mixtures--systems with almost indistinguishable microstructures but widely different dynamics. 
We have demonstrated that the theory is able to capture subtle differences in the static structure factors and amplify these to  account for the distinct LJ and WCA dynamics. 
Since the theory only uses $S^{\alpha\beta}(k)$ as input, all the relevant microstructural information is assumed to be fully encoded in the pair-correlations--a result that is consistent with recent machine-learning studies on these systems~\cite{Landes2020,paret2020}.
Moreover, owing to the improved predictive power of higher order GMCT compared to standard MCT, we have argued that our theory is also able to better understand the role of attraction in dense supercooled liquids.
We have also shown that highly polydisperse systems require a multi-component theory to properly describe the structure-dynamics link in supercooled liquids; this is because the single-component approximation ignores any   species-dependent structural correlations in $S(k)$, thus washing away many subtle but important features in the microstructure that subsequently compromise the predictive power of GMCT.

Lastly, we have illustrated that the systematic inclusion of more levels in the multi-component GMCT hierarchy yields quantitatively better predictions for the dynamics, at least based on the first few calculated GMCT levels. This gradual but systematic improvement is also consistent with earlier GMCT studies for single-component systems. In future work we will aim to push the boundaries of the highest level $n$ we can numerically solve~\cite{Luo2020_1,Luo2020_2}, in order to check whether the current GMCT framework might approach the exact scenario in the $n\xrightarrow[]{}\infty$ limit.
To conclude, we hope that our multi-component GMCT could be a useful tool to evaluate how static correlations influence the dynamics of supercooled liquids and to make reliable predictions about the dynamics of such liquids from static information only, thereby contributing to the final understanding of the glass transition.

\section{Acknowledgements}
We thank the Dutch Research Council (NWO) for financial support through a START-UP grant (C.L., V.E.D., and L.M.C.J.) and Vidi grant (L.M.C.J.).

\clearpage
\section{Supplementary Information}

\subsection{Derivation of multi-component GMCT}
\noindent In this section we present the full derivation of multi-component GMCT. We first introduce the Mori-Zwanzig formalism, which can be used to obtain exact dynamical equations for the correlation functions of arbitrary vectors of classical variables. Choosing the vector to consist of $1$-point (species dependent) density and current modes, we rederive  
standard multi-component MCT and highlight several technical details which will also prove to be useful for the derivation of GMCT. Finally, we generalize the vector to include multi-point density and current modes and retrieve the microscopic time-dependent multi-component GMCT equations.

\subsubsection{Mori-Zwanzig formalism}
\noindent Any dynamical vector $\bm{A}(t)$, whose elements $A_i(t)=[\bm{A}(t)]_i$ are functions of classical phase space variables, changes over time according to $\frac{d\bm{A}(t)}{dt}=\{\bm{A}(t),\mathcal{H}\}=i\mathcal{L}\bm{A}(t)$~\cite{zwanzig2001nonequilibrium}. Here, $\{\hdots\}$ denotes the Poisson bracket, $\mathcal{H}$ is the Hamiltonian of the system, and $\mathcal{L}$ depicts the classical Liouville operator. In analogy to vector algebra, one can define a projection operator $P$ to project onto the space spanned by $\bm{A}=\bm{A}(0)$:
\begin{eqnarray}
P=\sum_{ij} |A_i)T_{ij}(A_j|.
\end{eqnarray} 
Using the idempotent property $PP=P$, the matrix $\bm{T}$ can be determined via
\begin{eqnarray}
\sum_{j}T_{ij}(A_j|A_l)=\delta_{il},
\end{eqnarray}
where $(X|Y)=\langle X^*Y \rangle$ defines a scalar product of variables $X$ and $Y$,  $\langle\hdots\rangle$ is the ensemble average, and $\delta_{il}$ is the Kronecker delta function.
If we denote $\bm{G}=(\bm{A}|\bm{A})$ or $G_{ij}=[\bm{G}]_{ij}=(A_i|A_j)$,  then
\begin{equation}
    \bm{T}=\bm{G}^{-1}.
\end{equation}
Usually, though not necessarily, the elements in the vector $\bm{A}$ are chosen to be slow or quasi-conserved variables of the system. As a result, a projection only retains the slow part that is parallel to $\bm{A}$, and it removes the orthogonal or fast part, which can be obtained by using the complementary operator $1-P$.
Inserting several of those projections, the dynamical equation of $\bm{A}(t)$ can be written as \begin{equation}
    \frac{d\bm{A}(t)}{dt}=e^{i\mathcal{L}t}(P+1-P)i\mathcal{L}\bm{A}=e^{i\mathcal{L}t}Pi\mathcal{L}\bm{A}+e^{i\mathcal{L}t}(1-P)i\mathcal{L}\bm{A}=e^{i\mathcal{L}t}P\dot{\bm{A}}+e^{i\mathcal{L}t}(1-P)\dot{\bm{A}}
\end{equation}
If we then replace $e^{i\mathcal{L}t}$ in the last term by the identity~\cite{Reichman2005}
\begin{eqnarray}
e^{i\mathcal{L}t}=\int_0^td\tau e^{i\mathcal{L}(t-\tau)}iP\mathcal{L}e^{i(1-P)\mathcal{L}\tau}+e^{i(1-P)\mathcal{L}t},
\end{eqnarray}
we obtain
\begin{eqnarray}
|\dot{A}_k(t))=\sum_{ij}|A_i(t)) (\bm{G}^{-1})_{ij}(A_j|\dot{A}_k)-\int d\tau \sum_{ij}|A_i(t-\tau))(\bm{G}^{-1})_{ij}\left(f_j|f_k(\tau)\right)+|f_k(t)),
\end{eqnarray}
where we have introduced the so-called fluctuating force
\begin{eqnarray}
|f_k(t))\equiv e^{i(1-P)Lt}(1-P)|\dot{A}_k),
\end{eqnarray}
which, at $t=0$, evolves in time according to
\begin{eqnarray}
|f_k)=|\dot{A}_k)-\sum_{ij}|A_i)(\bm{G}^{-1})_{ij}(A_j|\dot{A}_k).
\end{eqnarray}
Physically, the fluctuating force represents the time evolution in the subspace orthogonal to $\bm{A}$, i.e., it constitutes the fast part of the dynamics. It should therefore be orthogonal to the slow variable, i.e.\ $(A_i|f_k(t))=0$, which is easily checked from the definition of $|f_k(t))$. Invoking this orthogonality, we may find the following dynamical equation for the correlation functions $C_{ij}(t)=(A_i|A_j(t))$,
\begin{equation}
\label{eq:C_general}
    \dot{C}_{ij}(t)=\sum_{lm}C_{il}(t) (\bm{G}^{-1})_{lm}(A_m|\dot{A}_j)-\int d\tau \sum_{lm}C_{il}(t-\tau)(\bm{G}^{-1})_{lm}(f_m|f_j(\tau)).
\end{equation}
Note that in this formalism, correlation functions are also written as scalar products.
We point out that the above equation is exact and that the main difficulty of calculating the correlation functions $C_{ij}(t)$ resides in finding an (approximate) expression for the memory function $K_{mj}(\tau)=(f_m|f_j(\tau))$. The procedure above, which effectively separates the dynamics of a system into a relevant (slow) and irrelevant (fast) part, is called the Mori-Zwanzig formalism and both MCT and GMCT are based on it.

\subsubsection{Multi-component MCT}
\noindent Now let us focus on a multi-component model system consisting of $N$ particles and $M$ species. For such a system we will choose as our slow variables
\begin{eqnarray}
\bm{A}=[\{\rho^{\alpha}_{\bm{k}}\}_{\alpha\in \{1,2,...,M\}}, \{j^{\alpha}_{\bm{k}}\}_{\alpha\in \{1,2,...,M\}}],
\end{eqnarray}
where $\rho^{\alpha}_{\bm{k}}=\sum_{i=1} ^{N_{\alpha}}e^{i\bm{k}\cdot\bm{r}_i}/\sqrt{N}$ is a density mode for species $\alpha$, $j^{\alpha}_{\bm{k}}=-i\dot{\rho}^{\alpha}_{\bm{k}}$ is the corresponding current mode, and the wavevector $\bm{k}$ probes the length scale of interest. The matrix $C_{ij}$ can then be regarded as four blocks, written schematically as 
$\begin{bmatrix}
C_{11}=(\rho|\rho) & C_{12}=(\rho|j)\\
C_{21}=(j|\rho) & C_{22}=(j|j)
\end{bmatrix}$,
although we will primarily focus on the dynamics of the upper-left block with elements $(\rho^{\alpha}_{\bm{k}}|\rho^{\beta}_{\bm{k}}(t))$.
Making use of the fact that
$(\rho^{\alpha}_{\bm{k}}|\rho^{\beta}_{\bm{k}})=S_{\alpha\beta}(k)$, $(\dot{\rho}^{\alpha}_{\bm{k}}|\dot{\rho}^{\beta}_{\bm{k}})=\delta_{\alpha\beta}k^2\frac{k_BT}{m_{\alpha}}x_{\alpha}\equiv J_{\alpha\beta}(k)$, and $(\rho^{\alpha}_{\bm{k}}|\dot{\rho}^{\beta}_{\bm{k}})=0$, it is easy to show that 
\begin{eqnarray}
\bm{G}=\begin{bmatrix}
\bm{S}(k) & \bm{0}\\
\bm{0} & \bm{J}(k)
\end{bmatrix},
\end{eqnarray}
and 
\begin{eqnarray}
(\bm{A}|\dot{\bm{A}})=\begin{bmatrix}
\bm{0} &i\bm{J}(k) \\
i\bm{J}(k) & \bm{0}
\end{bmatrix}.
\end{eqnarray}
with $\bm{S}$ and $\bm{J}$ depicting $M\times M$ matrices. Moreover, the fluctuating force can be calculated to give
\begin{eqnarray}
\bm{f}&&=\left[\left\{\dot{\rho}^{\alpha}_{\bm{k}}\right\}, \left\{\dot{j}^{\alpha}_{\bm{k}}\right\}\right]-\left[\{{\rho}^{\alpha}_{\bm{k}}\}, \{{j}^{\alpha}_{\bm{k}}\}\right]\begin{bmatrix}
\bm{S}^{-1}(k) & \bm{0}\\
\bm{0} & \bm{J}^{-1}(k)
\end{bmatrix}
\begin{bmatrix}
\bm{0} &i\bm{J}(k) \\
i\bm{J}(k) & \bm{0}
\end{bmatrix}
\\&&=\left[\{\dot{\rho}^{\alpha}_{\bm{k}}\}, \{\dot{j}^{\alpha}_{\bm{k}}\}\right]-\left[i\{{j}^{\alpha}_{\bm{k}}\}, i\{{\rho}^{\alpha}_{\bm{k}}\}\bm{S}^{-1}(k)\bm{J}(k)\right]
\\&&=\left[\bm{0},\{\dot{j}^{\alpha}_{\bm{k}}\}-i\{{\rho}^{\alpha}_{\bm{k}}\}\bm{S}^{-1}(k)\bm{J}(k)\right],
\end{eqnarray}
which implies that each element of $\bm{f}$ yields $f_k^{\alpha}=\dot{j}^{\alpha}_{\bm{k}}-i\sum_{\beta\gamma}{\rho}^{\beta}_{\bm{k}}\left({S}^{-1}(k)\right)_{\beta\gamma}{J}_{\gamma\alpha}(k)$.
Using these ingredients allows us to write down a dynamical equation for $F_{\alpha\beta}(\bm{k},t)=(\rho^{\alpha}_{\bm{k}}|\rho^{\beta}_{\bm{k}}(t))$ in the following manner:
\begin{equation}\label{MCT_eom}
    \ddot{F}_{\alpha\beta}(\bm{k},t)+\sum_{\gamma\theta}{F}_{\alpha\gamma}(\bm{k},t)\left(\bm{S}^{-1}(k)\right)_{\gamma\theta}{J}_{\theta\beta}(k)+\int d\tau \dot{{F}}_{\alpha\gamma}(\bm{k},t-\tau)\left(\bm{J}^{-1}(k)\right)_{\gamma\theta}K_{\theta\beta}(\bm{k},\tau)=0.
\end{equation}

As stated previously, the memory function $K_{\theta\beta}(\bm{k},\tau)=(f_{\bm{k}}^{\theta}|f_{\bm{k}}^{\beta}(\tau))$ forms the main problem for any analytical progress. We therefore seek to approximate it by inserting the projection operator
\begin{equation}
\label{eq:P2}
    P_2=\sum^{\alpha\beta\bm{q}\bm{p}} _{\alpha'\beta'\bm{q'}\bm{p'}}|{\rho}^{\alpha}_{\bm{q}}{\rho}^{\beta}_{\bm{p}})T^{\alpha\beta\bm{q}\bm{p}}_{\alpha'\beta'\bm{q'}\bm{p'}}({\rho}^{\alpha'}_{\bm{q'}}{\rho}^{\beta'}_{\bm{p'}}|
\end{equation}
in front of the fluctuating force and replacing the projected time evolution by a full one, i.e.\  $(f_{\bm{k}}^{\theta}|f_{\bm{k}}^{\beta}(\tau))\approx (f_{\bm{k}}^{\theta}P_2|e^{iLt}|P_2 f_{\bm{k}}^{\beta})$.
The first part of this approximation, i.e.\ applying the projector $P_2$, is rooted in the assumption that the dominant contributions to $f_{\bm{k}}^{\theta}$ arise from slow pair-density modes $\rho^{\alpha}_{\bm{q}}\rho^{\beta}_{\bm{p}}$. The replacement of the projected time evolution operator $e^{i(1-P)Lt}$ in the memory function by $e^{iLt}$ is mainly to keep calculations tractable. Note that we have introduced the most general definition of $P_2$ in which all wavevectors and species of the involved density modes can be different. However, such a strict definition will prove not to be necessary, since the wavevectors are constrained due to translational invariance of the system and a coupling to the wavevector $\bm{k}$ in $f_{\bm{k}}^{\theta}$~\cite{Reichman2005}. How this simplifies $P_2$ will become more apparent in the following parts of the derivation of MCT.
We proceed by first specifying the normalization of $P_2$. Using the property $P_2P_2=P_2$ gives
\begin{eqnarray}
\sum_{\alpha'\beta'\bm{q'}\bm{p'}}T^{\alpha\beta\bm{q}\bm{p}}_{\alpha'\beta'\bm{q'}\bm{p'}}({\rho}^{\alpha'}_{\bm{q'}}{\rho}^{\beta'}_{\bm{p'}}|{\rho}^{\gamma}_{\bm{m}}{\rho}^{\theta}_{\bm{n}})=\delta_{\alpha\gamma}\delta_{\beta\theta}\delta_{\bm{qm}}\delta_{\bm{pn}},
\end{eqnarray}
where, under the assumption of Gaussian factorization, we have
\begin{eqnarray}
({\rho}^{\alpha'}_{\bm{q'}}{\rho}^{\beta'}_{\bm{p'}}|{\rho}^{\gamma}_{\bm{m}}{\rho}^{\theta}_{\bm{n}})\approx
S^{\alpha'\gamma}_{m}\delta_{q'm}S^{\beta'\theta}_{n}\delta_{p'n}+S^{\alpha'\theta}_{n}\delta_{q'n}S^{\beta'\gamma}_{m}\delta_{p'm},
\end{eqnarray}
such that
\begin{eqnarray}
\sum_{\alpha'\beta'}T^{\alpha\beta\bm{q}\bm{p}}_{\alpha'\beta'\bm{m}\bm{n}}S^{\alpha'\gamma}_{m}S^{\beta'\theta}_{n}
+\sum_{\alpha'\beta'}T^{\alpha\beta\bm{q}\bm{p}}_{\alpha'\beta'\bm{n}\bm{m}}S^{\alpha'\theta}_{n}S^{\beta'\gamma}_{m}
=\delta_{\alpha\gamma}\delta_{\beta\theta}\delta_{\bm{qm}}\delta_{\bm{pn}}.
\end{eqnarray}
Then, making use of the fact that $\alpha'$ and $\beta'$ in the second term on the left-hand side are interchangeable and that $T^{\alpha\beta\bm{q}\bm{p}}_{\alpha'\beta'\bm{m}\bm{n}}
=T^{\alpha\beta\bm{q}\bm{p}}_{\beta'\alpha'\bm{n}\bm{m}}$ by symmetry,
we are allowed to rewrite 
\begin{eqnarray}
\label{eq:T_2}
2\sum_{\alpha'\beta'}T^{\alpha\beta\bm{q}\bm{p}}_{\alpha'\beta'\bm{m}\bm{n}}
S^{\alpha'\gamma}_{m}S^{\beta'\theta}_{n}
=\delta_{\alpha\gamma}\delta_{\beta\theta}\delta_{\bm{qm}}\delta_{\bm{pn}},
\end{eqnarray}
which defines our normalization tensor $T^{\alpha\beta\bm{q}\bm{p}}_{\alpha'\beta'\bm{m}\bm{n}}$.
Having fully specified our operator, we now seek to calculate the projected force $P_2|f_{\bm{k}}^{\mu})=P_2\left[|\dot{j}^{\mu}_{\bm{k}})-i\sum_{\theta\phi}|{\rho}^{\theta}_{\bm{k}})\left({S}^{-1}(k)\right)_{\theta\phi}{J}_{\phi\mu}(k)\right]$.
In particular, we have
\begin{eqnarray}
\label{eq:part1a}
&&({\rho}^{\alpha'}_{\bm{q'}}{\rho}^{\beta'}_{\bm{p'}}|\dot{j}^{\mu}_{\bm{k}})=-(\dot{\rho}^{\alpha'}_{\bm{q'}}{\rho}^{\beta'}_{\bm{p'}}|{j}^{\mu}_{\bm{k}})-({\rho}^{\alpha'}_{\bm{q'}}\dot{\rho}^{\beta'}_{\bm{p'}}|{j}^{\mu}_{\bm{k}}) \nonumber \\
&&=i \frac{1}{\sqrt{N}}\frac{k_BT}{m_\mu}\delta_{\bm{q'},\bm{k-p'}}\left[\bm{q'}\cdot\bm{k}S_{p'}^{\alpha'\beta'}\delta_{\mu\alpha'}+\bm{p'}\cdot\bm{k}S_{q'}^{\alpha'\beta'}\delta_{\mu\beta'}\right],
\end{eqnarray}
and, using the equation from the generalized Ornstein-Zernike relation that factorizes static triplet correlations into pair correlations with the direct triplet correlation functions $c_3$ as corrections~\cite{barrat1988,sciortino2001debye},
\begin{eqnarray}
\label{eq:S3}
S_3^{\alpha'\beta'\theta}(\bm{q}',\bm{p}',\bm{k}')=({\rho}^{\alpha'}_{\bm{q'}}{\rho}^{\beta'}_{\bm{p'}}|{\rho}^{\theta}_{\bm{k}})=\delta_{\bm{q'+p'},\bm{k}}\frac{1}{\sqrt{N}}\sum_{\epsilon\sigma\eta}S^{\alpha'\epsilon}_{q'}S^{\beta'\sigma}_{p'}S^{\theta\eta}_{k}\left[\delta_{\epsilon\sigma}\delta_{\sigma\eta}\delta_{\epsilon\eta}/x_{\epsilon}^2+{\rho}^2c_3^{\epsilon\sigma\eta}(\bm{q'},\bm{p'})\right],
\end{eqnarray}
we may also obtain
\begin{eqnarray}
\label{eq:part2a}
&&-i({\rho}^{\alpha'}_{\bm{q'}}{\rho}^{\beta'}_{\bm{p'}}|\sum_{\theta\phi}|{\rho}^{\theta}_{\bm{k}})\left({S}^{-1}(k)\right)_{\theta\phi}{J}_{\phi\mu}(k))
\nonumber\\
&&=-i\delta_{\bm{q'+p'},\bm{k}}\frac{1}{\sqrt{N}}\frac{k_BT}{m_\mu}\left[S^{\alpha'\mu}_{q'}S^{\beta'\mu}_{p'} k^2/x_{\mu}+\sum_{\epsilon\sigma}S^{\alpha'\epsilon}_{q'}S^{\beta'\sigma}_{p'} {\rho}^2c_3^{\epsilon\sigma\mu}(\bm{q'},\bm{p'})k^2x_{\mu}\right].
\end{eqnarray}
Combining Eqs.~(\ref{eq:part1a}) and (\ref{eq:part2a}), we arrive at
\begin{eqnarray}\label{proj_f1}
&&({\rho}^{\alpha'}_{\bm{q'}}{\rho}^{\beta'}_{\bm{p'}}|f_{\bm{k}}^{\mu})
\nonumber\\
&&=-i\delta_{\bm{q'+p'},\bm{k}}\frac{{\rho}}{\sqrt{N}}\frac{k_BT}{m_\mu}
\left[\bm{q'}\cdot\bm{k}S_{p'}^{\beta'\mu}\sum_{\theta}S_{q'}^{\alpha'\theta}c_{q'}^{\theta\mu}
+\bm{p'}\cdot\bm{k}S_{q'}^{\alpha'\mu}\sum_{\theta}S_{p'}^{\beta'\theta}c_{p'}^{\theta\mu}
+\sum_{\epsilon\sigma}S^{\alpha'\epsilon}_{q'}S^{\beta'\sigma}_{p'} \rho c_3^{\epsilon\sigma\mu}(\bm{q'},\bm{p'})k^2x_{\mu}\right],
\end{eqnarray}
where $c_{q}^{\alpha\beta}=(\delta_{\alpha\beta}/x_{\alpha}-\left(\bm{S}^{-1}(q)\right)_{\alpha\beta})/{\rho}$ is the direct correlation function.
An inspection of Eq.~(\ref{proj_f1}) now clearly shows the wavevector constraint on the pair density modes produced by projecting on $|f_{\bm{k}}^{\mu})$, i.e.\ only terms $({\rho}^{\alpha'}_{\bm{q'}}{\rho}^{\beta'}_{\bm{k-q'}}| $ for a given $\bm{k}$ contribute to the projected fluctuating force. In other words, we can equally define the projection operator as $P_2=\sum^{\alpha\beta\bm{q}} _{\alpha'\beta'\bm{q'}}|{\rho}^{\alpha}_{\bm{q}}{\rho}^{\beta}_{\bm{k-q}})T^{\alpha\beta\bm{q}}_{\alpha'\beta'\bm{q'}}({\rho}^{\alpha'}_{\bm{q'}}{\rho}^{\beta'}_{\bm{k-q'}}|$ instead of the most general one [Eq.~(\ref{eq:P2})]. Utilizing our results we finally obtain for the full projection
\begin{eqnarray}
P_2|f_{\bm{k}}^{\mu})&&=\sum^{\alpha\beta\bm{q}\bm{p}} _{\alpha'\beta'\bm{q'}\bm{p'}}|{\rho}^{\alpha}_{\bm{q}}{\rho}^{\beta}_{\bm{p}})T^{\alpha\beta\bm{q}\bm{p}}_{\alpha'\beta'\bm{q'}\bm{p'}}({\rho}^{\alpha'}_{\bm{q'}}{\rho}^{\beta'}_{\bm{p'}}|f_{\bm{k}}^{\mu})\\
&&=-i\sum^{\alpha\beta\bm{q}\bm{p}} _{\alpha'\beta'\bm{q'}\bm{p'}}|{\rho}^{\alpha}_{\bm{q}}{\rho}^{\beta}_{\bm{p}})T^{\alpha\beta\bm{q}\bm{p}}_{\alpha'\beta'\bm{q'}\bm{p'}}
\delta_{\bm{q'+p'},\bm{k}}\frac{{\rho}}{\sqrt{N}}\frac{k_BT}{m_\mu}
\nonumber\\
&&\times\left[\bm{q'}\cdot\bm{k}S_{p'}^{\beta'\mu}\sum_{\theta}S_{q'}^{\alpha'\theta}c_{q'}^{\theta\mu}
+\bm{p'}\cdot\bm{k}S_{q'}^{\alpha'\mu}\sum_{\theta}S_{p'}^{\beta'\theta}c_{p'}^{\theta\mu}
+\sum_{\epsilon\sigma}S^{\alpha'\epsilon}_{q'}S^{\beta'\sigma}_{p'} {\rho}c_3^{\epsilon\sigma\mu}(\bm{q'},\bm{p'})k^2x_{\mu}\right]
\nonumber\\
&&=-\frac{i}{2}\sum^{\alpha\beta\bm{q}\bm{p}} _{\bm{q'}\bm{p'}}|{\rho}^{\alpha}_{\bm{q}}{\rho}^{\beta}_{\bm{p}})
\delta_{\bm{q'+p'},\bm{k}}\frac{{\rho}}{\sqrt{N}}\frac{k_BT}{m_\mu}
\nonumber\\
&&\times\left[\bm{q'}\cdot\bm{k}\sum_{\theta}c_{q'}^{\theta\mu}\delta_{\alpha\theta}\delta_{\beta\mu}\delta_{\bm{qq'}}\delta_{\bm{pp'}}
+\bm{p'}\cdot\bm{k}\sum_{\theta}c_{p'}^{\theta\mu}\delta_{\alpha\mu}\delta_{\beta\theta}\delta_{\bm{qq'}}\delta_{\bm{pp'}}
+\sum_{\epsilon\sigma} {\rho}c_3^{\epsilon\sigma\mu}(\bm{q'},\bm{p'})k^2x_{\mu}\delta_{\alpha\epsilon}\delta_{\beta\sigma}\delta_{\bm{qq'}}\delta_{\bm{pp'}}\right]
\nonumber\\
&&=-\frac{i}{2}\sum_{\alpha\beta\bm{q}} |{\rho}^{\alpha}_{\bm{q}}{\rho}^{\beta}_{\bm{k-q}})
\frac{{\rho}}{\sqrt{N}}\frac{k_BT}{m_\mu}
\left[\bm{q}\cdot\bm{k}c_{q}^{\alpha\mu}\delta_{\beta\mu}
+(\bm{k-q})\cdot\bm{k}c_{|\bm{k-q}|}^{\beta\mu}\delta_{\alpha\mu}
+ {\rho}k^2x_{\mu}c_3^{\alpha\beta\mu}(\bm{q},\bm{k-q})\right],
\end{eqnarray}
and, consequently, the memory function yields
\begin{eqnarray}
\label{eq:Mmct}
(f_{\bm{k}}^{\mu}|f_{\bm{k}}^{\nu}(t))=\frac{{\rho}^2}{4N}\sum_{\alpha\beta\bm{q}}\sum_{\alpha'\beta'\bm{q'}}\frac{k_BT}{m_\mu} \mathcal{V}_{\alpha\beta\mu}(\bm{q},\bm{k-q},\bm{k})({\rho}^{\alpha}_{\bm{q}}{\rho}^{\beta}_{\bm{k-q}}|{\rho}^{\alpha'}_{\bm{q'}}(t){\rho}^{\beta'}_{\bm{k-q'}}(t))\mathcal{V}_{\alpha'\beta'\nu}(\bm{q'},\bm{k-q'},\bm{k})\frac{k_BT}{m_\nu},
\end{eqnarray}
with the vertex, which represents the coupling strength between different wavevectors, given by
\begin{eqnarray}
\label{eq:vertex}
\mathcal{V}_{\alpha\beta\mu}(\bm{q},\bm{k-q},\bm{k})=\bm{q}\cdot\bm{k}c_{q}^{\alpha\mu}\delta_{\beta\mu}
+(\bm{k-q})\cdot\bm{k}c_{|\bm{k-q}|}^{\beta\mu}\delta_{\alpha\mu}
+ {\rho}k^2x_{\mu}c_3^{\alpha\beta\mu}(\bm{q},\bm{k-q}).
\end{eqnarray}
Note that, apart from the unknown dynamic density correlation functions $({\rho}^{\alpha}_{\bm{q}}{\rho}^{\beta}_{\bm{k-q}}|{\rho}^{\alpha'}_{\bm{q'}}(t){\rho}^{\beta'}_{\bm{k-q'}}(t))$, all other terms in the memory function [Eq.\ (\ref{eq:Mmct})] are static, and in principle known, equilibrium properties of the system. 

In standard MCT, the multi-point density correlation functions $({\rho}^{\alpha}_{\bm{q}}{\rho}^{\beta}_{\bm{k-q}}|{\rho}^{\alpha'}_{\bm{q'}}(t){\rho}^{\beta'}_{\bm{k-q'}}(t))$ are simplified in two steps. The first one is the so-called diagonal approximation, i.e.\ 
\begin{eqnarray}
   &&({\rho}^{\alpha}_{\bm{q}}{\rho}^{\beta}_{\bm{k-q}}|{\rho}^{\alpha'}_{\bm{q'}}(t){\rho}^{\beta'}_{\bm{k-q'}}(t))\approx  ({\rho}^{\alpha}_{\bm{q}}{\rho}^{\beta}_{\bm{k-q}}|{\rho}^{\alpha'}_{\bm{q'}}(t){\rho}^{\beta'}_{\bm{k-q'}}(t))(\delta_{\bm{q,q'}}+\delta_{\bm{q,k-q'}})
   \nonumber\\
   &&=({\rho}^{\alpha}_{\bm{q}}{\rho}^{\beta}_{\bm{k-q}}|{\rho}^{\alpha'}_{\bm{q}}(t){\rho}^{\beta'}_{\bm{k-q}}(t))\delta_{\bm{q,q'}}+({\rho}^{\alpha}_{\bm{q}}{\rho}^{\beta}_{\bm{k-q}}|{\rho}^{\alpha'}_{\bm{k-q}}(t){\rho}^{\beta'}_{\bm{q}}(t))\delta_{\bm{q,k-q'}},
\end{eqnarray}
which, using $\mathcal{V}_{\beta'\alpha'\nu}(\bm{k-q},\bm{q},\bm{k})=\mathcal{V}_{\alpha'\beta'\nu}(\bm{q},\bm{k-q},\bm{k})$, implies that
\begin{eqnarray}
\label{eq:part1}
(f_{\bm{k}}^{\mu}|f_{\bm{k}}^{\nu}(t))=\frac{{\rho}^2}{4N}\sum_{\alpha\beta\bm{q}}\sum_{\alpha'\beta'}\frac{k_BT}{m_\mu} \mathcal{V}_{\alpha\beta\mu}(\bm{q},\bm{k-q},\bm{k})({\rho}^{\alpha}_{\bm{q}}{\rho}^{\beta}_{\bm{k-q}}|{\rho}^{\alpha'}_{\bm{q}}(t){\rho}^{\beta'}_{\bm{k-q}}(t))\mathcal{V}_{\alpha'\beta'\nu}(\bm{q},\bm{k-q},\bm{k})\frac{k_BT}{m_\nu}
\nonumber\\
+\frac{{\rho}^2}{4N}\sum_{\alpha\beta\bm{q}}\sum_{\alpha'\beta'}\frac{k_BT}{m_\mu} \mathcal{V}_{\alpha\beta\mu}(\bm{q},\bm{k-q},\bm{k})({\rho}^{\alpha}_{\bm{q}}{\rho}^{\beta}_{\bm{k-q}}|{\rho}^{\alpha'}_{\bm{k-q}}(t){\rho}^{\beta'}_{\bm{q}}(t))\mathcal{V}_{\alpha'\beta'\nu}(\bm{k-q},\bm{q},\bm{k})\frac{k_BT}{m_\nu}
\\=\frac{{\rho}^2}{2N}\sum_{\alpha\beta\bm{q}}\sum_{\alpha'\beta'}\frac{k_BT}{m_\mu} \mathcal{V}_{\alpha\beta\mu}(\bm{q},\bm{k-q},\bm{k})({\rho}^{\alpha}_{\bm{q}}{\rho}^{\beta}_{\bm{k-q}}|{\rho}^{\alpha'}_{\bm{q}}(t){\rho}^{\beta'}_{\bm{k-q}}(t))\mathcal{V}_{\alpha'\beta'\nu}(\bm{q},\bm{k-q},\bm{k})\frac{k_BT}{m_\nu}.
\end{eqnarray}
The second step is Gaussian factorization of the multi-point density correlations, 
\begin{equation}
\label{eq:appr}
({\rho}^{\alpha}_{\bm{q}}{\rho}^{\beta}_{\bm{k-q}}|{\rho}^{\alpha'}_{\bm{q}}(t){\rho}^{\beta'}_{\bm{k-q}}(t))\approx ({\rho}^{\alpha}_{\bm{q}}|{\rho}^{\alpha'}_{\bm{q}}(t))({\rho}^{\beta}_{\bm{k-q}}|{\rho}^{\beta'}_{\bm{k-q}}(t))=F_{\alpha\alpha'}(\bm{q},t)F_{\beta\beta'}(\bm{k-q},t),
\end{equation}
so that the memory function in standard MCT becomes a function of the $2$-point density correlation function,
\begin{eqnarray}
(f_{\bm{k}}^{\mu}|f_{\bm{k}}^{\nu}(t))=\frac{{\rho}^2}{2N}\sum_{\bm{q}}\sum_{\alpha\beta\alpha'\beta'}\frac{k_BT}{m_\mu} \mathcal{V}_{\alpha\beta\mu}(\bm{q},\bm{k-q},\bm{k})F_{\alpha\alpha'}(\bm{q},t)F_{\beta\beta'}(\bm{k-q},t)\mathcal{V}_{\alpha'\beta'\nu}(\bm{q'},\bm{k-q'},\bm{k})\frac{k_BT}{m_\nu}
\nonumber\\=\frac{{\rho}}{2}\int \frac{\bm{dq}}{(2\pi)^3}\sum_{\alpha\beta\alpha'\beta'}\frac{k_BT}{m_\mu} \mathcal{V}_{\alpha\beta\mu}(\bm{q},\bm{k-q},\bm{k})F_{\alpha\alpha'}(\bm{q},t)F_{\beta\beta'}(\bm{k-q},t)\mathcal{V}_{\alpha'\beta'\nu}(\bm{q'},\bm{k-q'},\bm{k})\frac{k_BT}{m_\nu},
\end{eqnarray}
where in the last step we have assumed the thermodynamic limit to write $\sum_{\bm{q}}=\frac{V}{(2\pi)^3}\int d \bm{q} $.
Overall, the above set of approximations renders Eq.~(\ref{MCT_eom}) a closed equation for $F_{\alpha\beta}(\bm{k},t)$ which can be solved self-consistently. 

\subsubsection{Multi-component GMCT}
\noindent The main idea of GMCT is to avoid the most severe approximation of MCT, namely the factorization approximation of Eq.~(\ref{eq:appr}), and instead use an explicit expression for the diagonal 4-point density correlation functions $({\rho}^{\alpha}_{\bm{q}}{\rho}^{\beta}_{\bm{k-q}}|{\rho}^{\alpha'}_{\bm{q}}(t){\rho}^{\beta'}_{\bm{k-q}}(t))$. More generally, we seek to retrieve the exact equations of motion for arbitrary diagonal $2n$-point density correlation functions $(\rho^{\alpha_1}_{\bm{k_1}}\rho^{\alpha_2}_{\bm{k_2}}\hdots \rho^{\alpha_n}_{\bm{k_n}}|\rho^{\beta_1}_{\bm{k_1}}(t)\rho^{\beta_2}_{\bm{k_2}}(t)\hdots \rho^{\beta_n}_{\bm{k_n}}(t))$, and develop a hierarchy of equations relating them to each other. Following previous work on single component systems~\cite{Janssen2015a}, we generalize the vector $\bm{A}$ to include the $n$-point density modes and associated current modes at a given series of wavevectors $\{\bm{k}_1,\bm{k}_2,\hdots,\bm{k}_n\}$ (we assume $\bm{k}_i\neq \bm{k}_j$ for $i\neq j$),
\begin{eqnarray}
\bm{A}=\left[\left\{\rho^{\alpha_1}_{\bm{k_1}}\rho^{\alpha_2}_{\bm{k_2}}\hdots \rho^{\alpha_n}_{\bm{k_n}}\right\},-i\left\{ \frac{d}{dt} (\rho^{\alpha_1}_{\bm{k_1}}\rho^{\alpha_2}_{\bm{k_2}}\hdots \rho^{\alpha_n}_{\bm{k_n}})\right\}\right]\equiv [\bm{A}_1, \bm{A}_2],
\end{eqnarray}
where we point out that there are $M^n$ elements in both $\bm{A}_1$ and $\bm{A}_2$.
Employing the same procedure as in the derivation of MCT and realizing that $\bm{A}_2=-i\dot{\bm{A}}_1$ and $(\bm{A}_1|\bm{A}_2)=0$, we can calculate generalized versions of $\bm{G}$, $(\bm{A}|\bm{\dot{A}})$, and $\bm{f}$ in Eq.~(\ref{eq:C_general}). In particular, we have (adding the superscript $(n)$ to distinguish between different levels with $n=1$ referring to MCT),
\begin{eqnarray}
\bm{G}^{(n)}=
\begin{bmatrix}
\bm{S}^{(n)}(\{k_i\}) & \bm{0}\\
\bm{0} & \bm{J}^{(n)}(\{k_i\})
\end{bmatrix},
\end{eqnarray}
and 
\begin{eqnarray}
(\bm{A}|\dot{\bm{A}})=\begin{bmatrix}
\bm{0} &i\bm{J}^{(n)}(\{k_i\}) \\
i\bm{J}^{(n)}(\{k_i\}) & \bm{0}
\end{bmatrix}.
\end{eqnarray}
Here, $\bm{S}^{(n)}$ and $\bm{J}^{(n)}$ are $M^n\times M^n$ matrices with elements
\begin{eqnarray}
&&{S}_{\{\alpha_i\} ;\{\beta_i\}}^{(n)} (\{k_i\})=(\rho^{\alpha_1}_{\bm{k_1}}\rho^{\alpha_2}_{\bm{k_2}}\hdots \rho^{\alpha_n}_{\bm{k_n}}|\rho^{\beta_1}_{\bm{k_1}}\rho^{\beta_2}_{\bm{k_2}}\hdots \rho^{\beta_n}_{\bm{k_n}})
\nonumber\\
&&\approx (\rho^{\alpha_1}_{\bm{k_1}}|\rho^{\beta_1}_{\bm{k_1}})
(\rho^{\alpha_2}_{\bm{k_2}}|\rho^{\beta_2}_{\bm{k_2}})\hdots (\rho^{\alpha_n}_{\bm{k_n}}|\rho^{\beta_n}_{\bm{k_n}})
\nonumber\\
&&=S^{\alpha_1\beta_1}_{k_1}S^{\alpha_2\beta_2}_{k_2}\hdots S^{\alpha_n\beta_n}_{k_n}=\prod_{i=1}^{n}S^{\alpha_i\beta_i}({k_i}),
\end{eqnarray}
and 
\begin{eqnarray}
&&{J}_{\{\alpha_i\} ;\{\beta_i\}}^{(n)} (\{k_i\})=
\left( \frac{d}{dt} (\rho^{\alpha_1}_{\bm{k_1}}\rho^{\alpha_2}_{\bm{k_2}}\hdots \rho^{\alpha_n}_{\bm{k_n}}) \middle|
\frac{d}{dt} 
(\rho^{\beta_1}_{\bm{k_1}}\rho^{\beta_2}_{\bm{k_2}}\hdots \rho^{\beta_n}_{\bm{k_n}}) \right)
\nonumber\\
&&
=(\dot{\rho}^{\alpha_1}_{\bm{k_1}}\rho^{\alpha_2}_{\bm{k_2}}\hdots \rho^{\alpha_n}_{\bm{k_n}}|\dot{\rho}^{\beta_1}_{\bm{k_1}}\rho^{\beta_2}_{\bm{k_2}}\hdots \rho^{\beta_n}_{\bm{k_n}})+\hdots+(\rho^{\alpha_1}_{\bm{k_1}}\rho^{\alpha_2}_{\bm{k_2}}\hdots \dot{\rho}^{\alpha_n}_{\bm{k_n}}|\rho^{\beta_1}_{\bm{k_1}}\rho^{\beta_2}_{\bm{k_2}}\hdots \dot{\rho}^{\beta_n}_{\bm{k_n}})
\nonumber\\
&&
 \approx(\dot{\rho}^{\alpha_1}_{\bm{k_1}}|\dot{\rho}^{\beta_1}_{\bm{k_1}})
(\rho^{\alpha_2}_{\bm{k_2}}|\rho^{\beta_2}_{\bm{k_2}})\hdots (\rho^{\alpha_n}_{\bm{k_n}}|\rho^{\beta_n}_{\bm{k_n}})+\hdots+ (\rho^{\alpha_1}_{\bm{k_1}}|\rho^{\beta_1}_{\bm{k_1}})
(\rho^{\alpha_2}_{\bm{k_2}}|\rho^{\beta_2}_{\bm{k_2}})\hdots (\dot{\rho}^{\alpha_n}_{\bm{k_n}}|\dot{\rho}^{\beta_n}_{\bm{k_n}})
\nonumber\\
&&=\delta_{\alpha_1\beta_1}k_1^2\frac{k_BT}{m_{\alpha_1}}x_{\alpha_1}S^{\alpha_2\beta_2}_{k_2}\hdots S^{\alpha_n\beta_n}_{k_n}+\hdots + S^{\alpha_1\beta_1}_{k_1}S^{\alpha_2\beta_2}_{k_2} \hdots\delta_{\alpha_n\beta_n}k_{n}^2\frac{k_BT}{m_{\alpha_n}}x_{\alpha_n}
\nonumber\\
&&
=\sum_{i=1}^{n}\delta_{\alpha_i,\beta_i}\frac{k_BTx_{\alpha_i}k_i^2}{m_{\alpha_i}S^{\alpha_i\beta_i}_{k_i}}\prod_{j=1}^{n}S^{\alpha_j\beta_j}_{k_j},
\end{eqnarray}
respectively. Moreover, the generalized fluctuating force is given by
\begin{eqnarray}
\bm{f}&&=\left[\dot{\bm{A}}_1, \dot{\bm{A}}_2\right]-\left[\bm{A}_1, \bm{A}_2\right]\begin{bmatrix}
\bm{S}^{-1}(q) & \bm{0}\\
\bm{0} & \bm{J}^{-1}(q)
\end{bmatrix}
\begin{bmatrix}
\bm{0} &i\bm{J}(q) \\
i\bm{J}(q) & \bm{0}
\end{bmatrix}
\\&&=\left[\dot{\bm{A}}_1, \dot{\bm{A}}_2\right]-\left[i\bm{A}_2, i\bm{A}_1\bm{S}^{-1}(q)\bm{J}(q)\right]
\\&&=\left[\bm{0},\dot{\bm{A}}_2-i\bm{A}_1\bm{S}^{-1}(q)\bm{J}(q)\right],
\end{eqnarray}
so that its elements are,
\begin{eqnarray}
f^{(n)}_{\{\alpha_i\}}(\{k_i\})=-i\frac{d^2(\rho^{\alpha_1}_{\bm{k_1}}\rho^{\alpha_2}_{\bm{k_2}}\hdots \rho^{\alpha_n}_{\bm{k_n}})}{dt^2}-i\sum_{\{\beta_i\}\{\gamma_i\}}\rho^{\beta_1}_{\bm{k_1}}\rho^{\beta_2}_{\bm{k_2}}\hdots \rho^{\beta_n}_{\bm{k_n}}\left({S^{(n)}}\left({\{k_i\}}\right)\right)^{-1}_{\{\beta_i\};\{\gamma_i\}}{J}^{(n)}_{\{\gamma_i\}\{\alpha_i\}}\left(\{k_i\}\right).
\end{eqnarray}
Using these terms and the Mori-Zwanzig formalism we can determine the dynamical equation for arbitrary $2n$-point density correlation functions, which yields
\begin{align}
\ddot{F}^{(n)}_{\{\alpha_i\};\{\beta_i\}}(\{k_i\},t)&+\sum_{\{\gamma_i\}\{\theta_i\}}F^{(n)}_{\{\alpha_i\};\{\gamma_i\}}(\{k_i\},t)\left(S^{(n)}\right)^{-1}_{\{\gamma_i\};\{\theta_i\}}(\{k_i\})J^{(n)}_{\{\theta_i\};\{\beta_i\}}(\{k_i\})
\nonumber\\&
+\int_0^t d\tau \sum_{\{\gamma_i\}\{\theta_i\}}\dot{F}^{(n)}_{\{\alpha_i\};\{\gamma_i\}}(\{k_i\},t-\tau) \left(J^{(n)}\right)^{-1}_{\{\gamma_i\};\{\theta_i\}}(\{k_i\})K^{(n)}_{\{\theta_i\};\{\beta_i\}}(\{k_i\},\tau)=0.
\end{align}
This serves as the starting point in our main text (see Sec.~II). The memory kernel is again a correlation between (generalized) fluctuating forces 
$K^{(n)}_{\{\theta_i\};\{\beta_i\}}(\{k_i\},\tau)=
(f^{(n)}_{\{\theta_i\}}(\{k_i\})|f^{(n)}_{\{\beta_i\}}(\{k_i\},\tau))$ and we approximate it, in the same fashion as MCT, by projecting these forces onto $(n+1)$th order density modes and replacing the orthogonal time evolution with a full one, i.e.\  $(f^{(n)}_{\{\theta_i\}}(\{k_i\}) | f^{(n)}_{\{\beta_i\}}(\{k_i\},\tau))\approx (f^{(n)}_{\{\theta_i\}}(\{k_i\}) P^{(n)}|e^{iL\tau}P^{(n)} f^{(n)}_{\{\beta_i\}}(\{k_i\})).$
Inspired by single component GMCT~\cite{Janssen2015a} and the simplified shape of $P_2$ introduced for MCT, we define the $n$th order projection operator as the summation of orthogonal projection operators:
\begin{eqnarray}
P^{(n)}=\sum_{i}P^{(n)}_i 
\end{eqnarray}
with 
\begin{eqnarray}
P^{(n)}_i=\sum_{\bm{q'}\alpha'_0,\hdots,\alpha'_n}^{\bm{q},\alpha_0,\hdots,\alpha_n}|\rho^{\alpha_0}_{\bm{k_i-q}}\rho^{\alpha_1}_{\bm{k_1}}\rho^{\alpha_2}_{\bm{k_2}}\hdots\rho^{\alpha_{i-1}}_{\bm{k}_{i-1}}\rho^{\alpha_i}_{\bm{q}}\rho^{\alpha_{i+1}}_{\bm{k}_{i+1}}\hdots \rho^{\alpha_n}_{\bm{k_n}})
T_{\bm{q'}\alpha'_0,\hdots,\alpha'_n}^{\bm{q},\alpha_0,\hdots,\alpha_n}(i)
(\rho^{\alpha'_0}_{\bm{k_i-q'}}\rho^{\alpha'_1}_{\bm{k_1}}\rho^{\alpha'_2}_{\bm{k_2}}\hdots\rho^{\alpha'_{i-1}}_{\bm{k}_{i-1}}\rho^{\alpha'_i}_{\bm{q'}}\rho^{\alpha'_{i+1}}_{\bm{k}_{i+1}}\hdots \rho^{\alpha'_n}_{\bm{k_n}}|.
\end{eqnarray}
Note that in this notation $P_1^{(1)}=P_2$.
For $i\neq j$ and using Eq.~(\ref{eq:S3}), it can be checked that $P^{(n)}_iP^{(n)}_j\sim S_3(\bm{k}_i-\bm{q},\bm{q'},\bm{k}_i)S_3(\bm{q},\bm{k}_j,\bm{k}_j-\bm{q'}) \sim \frac{1}{N}\rightarrow 0$, thus demonstrating that $P^{(n)}_i$ is orthogonal to $P^{(n)}_j$.
The normalization $T_{\bm{q'}\alpha'_0,\hdots,\alpha'_n}^{\bm{q},\alpha_0,\hdots,\alpha_n}(i)$ is determined via the condition $P^{(n)}_iP^{(n)}_i=P^{(n)}_i$. This gives
\begin{eqnarray}
\sum_{q',\alpha'_0,\alpha'_1,\hdots,\alpha'_n}T_{\bm{q'}\alpha'_0,\hdots,\alpha'_n}^{\bm{q},\alpha_0,\hdots,\alpha_n}(i)(\rho^{\alpha'_0}_{\bm{k_i-q'}}\rho^{\alpha'_1}_{\bm{k_1}}\hdots\rho^{\alpha'_i}_{\bm{q'}}\hdots \rho^{\alpha'_n}_{\bm{k_n}}|\rho^{\beta_0}_{\bm{k_i-p}}\rho^{\beta_1}_{\bm{k_1}}\hdots\rho^{\beta_i}_{\bm{p}}\hdots \rho^{\beta_n}_{\bm{k_n}})=\delta_{\bm{q}\bm{p}}\delta_{\alpha_0\beta_0}\delta_{\alpha_1\beta_1}\hdots\delta_{\alpha_n\beta_n}
\end{eqnarray}
and hence
\begin{eqnarray}
&&\sum_{\bm{q'},\alpha'_0,\alpha'_1,\hdots,\alpha'_n}T_{\bm{q'}\alpha'_0,\hdots,\alpha'_n}^{\bm{q},\alpha_0,\hdots,\alpha_n}(i)
\nonumber\\
&&
\times\left[S^{\alpha'_0\beta_0}_{|\bm{k}_i-\bm{q'}|}S^{\alpha'_1\beta_1}_{k_1}\hdots S^{\alpha'_i\beta_i}_{q'}\hdots S^{\alpha'_n\beta_n}_{k_n}\delta_{\bm{q'p}}
+S^{\alpha'_0\beta_i}_{|\bm{k}_i-\bm{q'}|}S^{\alpha'_1\beta_1}_{k_1}\hdots S^{\alpha'_i\beta_0}_{q'}\hdots S^{\alpha'_n\beta_n}_{k_n}\delta_{\bm{p,k_i-q'}}\right]
=\delta_{\bm{q}\bm{q'}}\delta_{\alpha_0\beta_0}\delta_{\alpha_1\beta_1}\hdots\delta_{\alpha_n\beta_n},
\end{eqnarray}
which, using the symmetry property $(\rho^{\alpha'_0}_{\bm{k_i-q'}}\rho^{\alpha'_1}_{\bm{k_1}}\hdots\rho^{\alpha'_i}_{\bm{q'}}\hdots \hdots\rho^{\alpha'_n}_{\bm{k_n}}|=(\rho^{\alpha'_i}_{\bm{q'}}\rho^{\alpha'_1}_{\bm{k_1}}\hdots\rho^{\alpha'_0}_{\bm{k_i-q'}}\hdots \rho^{\alpha'_n}_{\bm{k_n}}|$, can be simplified to 
\begin{eqnarray}
\label{eq:TGMCT}
2\sum_{\alpha'_0,\alpha'_1,\hdots,\alpha'_n}T_{\bm{p},\alpha'_0,\hdots,\alpha'_n}^{\bm{q},\alpha_0,\hdots,\alpha_n}(i)
S^{\alpha\prime_0\beta_0}_{|\bm{k}_i-\bm{p}|}S^{\alpha\prime_1\beta_1}_{k_1}\hdots S^{\alpha\prime_i\beta_i}_{p}\hdots S^{\alpha\prime_n\beta_n}_{k_n}
=\delta_{\bm{q}\bm{p}}\delta_{\alpha_0\beta_0}\delta_{\alpha_1\beta_1}\hdots\delta_{\alpha_n\beta_n}.
\end{eqnarray}
This completes the characterization of the generalized projection operators $P^{(n)}_{j}$.

Now let us calculate the projected fluctuating force $P^{(n)}_j|f^{(n)}_{\{\mu_i\}}(\{k_i\}))$. In order to calculate the projection of the first term of $f^{(n)}_{\{\mu_i\}}(\{k_i\})$, we require an expression for
\begin{eqnarray}
\label{eq:rhof1}
&&-i(\rho^{\alpha'_0}_{\bm{k_j-q'}}\rho^{\alpha'_1}_{\bm{k_1}}\hdots\rho^{\alpha'_j}_{\bm{q'}}\hdots \rho^{\alpha'_n}_{\bm{k_n}}|\frac{d^2}{dt^2}(\rho^{\mu_1}_{\bm{k_1}}\rho^{\mu_2}_{\bm{k_2}}\hdots \rho^{\mu_n}_{\bm{k_n}}))\nonumber\\
&&=i\big(\frac{d}{dt}(\rho^{\alpha'_0}_{\bm{k_j-q'}}\rho^{\alpha'_1}_{\bm{k_1}}\hdots\rho^{\alpha'_j}_{\bm{q'}}\hdots \rho^{\alpha'_n}_{\bm{k_n}})|\frac{d}{dt}(\rho^{\mu_1}_{\bm{k_1}}\rho^{\mu_2}_{\bm{k_2}}\hdots \rho^{\mu_n}_{\bm{k_n}})\big)
\end{eqnarray}
Realizing that $(\dot{\rho}^{\alpha}_{\bm{k}}|{\rho}^{\beta}_{\bm{k}})=0$, we find only three non-zero contributions to this expression:
\begin{enumerate}
     \item $i(\dot{\rho}^{\alpha'_0}_{\bm{k_j-q'}}\rho^{\alpha'_1}_{\bm{k_1}}\hdots\rho^{\alpha'_j}_{\bm{q'}}\hdots \rho^{\alpha'_n}_{\bm{k_n}}|\rho^{\mu_1}_{\bm{k_1}}\rho^{\mu_2}_{\bm{k_2}}\hdots \dot{\rho}^{\mu_j}_{\bm{k_j}}\hdots \rho^{\mu_n}_{\bm{k_n}})
     \approx i\frac{1}{\sqrt{N}}\frac{k_BT}{m_{\mu_j}}(\bm{k}_j-\bm{q'})\cdot\bm{k}_j\delta_{\alpha'_0\mu_j}S_{\bm{q}'}^{\alpha'_j\mu_j}\prod_{i\neq j}S_{k_i}^{\alpha'_i\mu_i}$
     \item $i({\rho}^{\alpha'_0}_{\bm{k_j-q'}}\rho^{\alpha'_1}_{\bm{k_1}}\hdots\dot{\rho}^{\alpha'_j}_{\bm{q'}}\hdots \rho^{\alpha'_n}_{\bm{k_n}}|\rho^{\mu_1}_{\bm{k_1}}\rho^{\mu_2}_{\bm{k_2}}\hdots \dot{\rho}^{\mu_j}_{\bm{k_j}}\hdots \rho^{\mu_n}_{\bm{k_n}})
     \approx i\frac{1}{\sqrt{N}}\frac{k_BT}{m_{\mu_j}}\bm{q'}\cdot\bm{k}_j\delta_{\alpha'_j\mu_j}S_{|\bm{k}_j-\bm{q}'|}^{\alpha'_0\mu_j}\prod_{i\neq j}S_{k_i}^{\alpha'_i\mu_i}$
     \item for $i\neq j$,\\
     $i({\rho}^{\alpha'_0}_{\bm{k_j-q'}}\rho^{\alpha'_1}_{\bm{k_1}}\hdots\dot{\rho}^{\alpha'_i}_{\bm{k_i}}\hdots{\rho}^{\alpha'_j}_{\bm{q'}}\hdots \rho^{\alpha'_n}_{\bm{k_n}})|\rho^{\mu_1}_{\bm{k_1}}\rho^{\mu_2}_{\bm{k_2}}\hdots\dot{\rho}^{\mu'_i}_{\bm{k_i}}\hdots {\rho}^{\mu_j}_{\bm{k_j}}\hdots \rho^{\mu_n}_{\bm{k_n}})
     \approx i S^{\alpha_0'\alpha'_j\mu_j}_3(\bm{k}_j-\bm{q}',\bm{q}',\bm{k}_j)\delta_{\alpha'_i\mu_i}k_i^2\frac{k_BT}{m_{\mu_i}}x_{\mu_i} \prod_{l\neq i,l\neq j}S_{k_l}^{\alpha'_l\mu_l}$
\end{enumerate}
Thus, the right-hand side of Eq.~(\ref{eq:rhof1}) becomes
\begin{eqnarray}
\label{eq:rhof1r}
&&i\frac{1}{\sqrt{N}}\frac{k_BT}{m_{\mu_j}}\Big[(\bm{k}_j-\bm{q'})\cdot\bm{k}_j\delta_{\alpha'_0\mu_j}S_{\bm{q}'}^{\alpha'_j\mu_j}\prod_{i\neq j}S_{k_i}^{\alpha'_i\mu_i}
+\bm{q'}\cdot\bm{k}_j\delta_{\alpha'_j\mu_j}S_{|\bm{k}_j-\bm{q}'|}^{\alpha'_0\mu_j}\prod_{i\neq j}S_{k_i}^{\alpha'_i\mu_i}\Big]
\nonumber\\
&&+i\sum_{i\neq j}S^{\alpha_0'\alpha'_j\mu_j}_3(\bm{k}_j-\bm{q}',\bm{q}',\bm{k}_j)\delta_{\alpha'_i\mu_i}k_i^2\frac{k_BT}{m_{\mu_i}}x_{\mu_i} \prod_{l\neq i,l\neq j}S_{k_l}^{\alpha'_l\mu_l}
\end{eqnarray}
In comparison, for the projection of the second term of the projected force $f^{(n)}_{\{\mu_i\}}(\{k_i\})$
we need the term 
\begin{eqnarray}
&&-i({\rho}^{\alpha'_0}_{\bm{k_j-q'}}\rho^{\alpha'_1}_{\bm{k_1}}\hdots{\rho}^{\alpha'_j}_{\bm{q'}}\hdots \rho^{\alpha'_n}_{\bm{k_n}}|\sum_{\{\theta_i\}\{\gamma_i\}}\rho^{\theta_1}_{\bm{k_1}}\rho^{\theta_2}_{\bm{k_2}}\hdots \rho^{\theta_n}_{\bm{k_n}})\left({S^{(n)}}({\{k_i\}}))^{-1}\right)_{\{\theta_i\};\{\gamma_i\}}{J}^{(n)}_{\{\gamma_i\}\{\mu_i\}}(\{k_i\})\nonumber\\
&&=-i\sum_{\{\theta_i\}\{\gamma_i\}}S^{\alpha_0'\alpha'_j\theta_j}_3(\bm{k}_j-\bm{q}',\bm{q}',\bm{k}_j)\prod_{i\neq j}S_{k_i}^{\alpha'_i\theta_i}\left(\prod_{i=1}^{n}S^{\theta_i\gamma_i}_{k_i}\right)^{-1}
\sum_{i=1}^{n}\delta_{\gamma_i,\mu_i}\frac{k_BTx_{\gamma_i}k_i^2}{m_{\gamma_i}S^{\gamma_i\mu_i}_{k_i}}\prod_{j=1}^{n}S^{\gamma_j\mu_j}_{k_j},
\end{eqnarray}
where we have applied Gaussian factorization to write
\begin{eqnarray}
({\rho}^{\alpha'_0}_{\bm{k_j-q'}}\rho^{\alpha'_1}_{\bm{k_1}}\hdots{\rho}^{\alpha'_j}_{\bm{q'}}\hdots \rho^{\alpha'_n}_{\bm{k_n}}|\rho^{\theta_1}_{\bm{k_1}}\rho^{\theta_2}_{\bm{k_2}}\hdots \rho^{\theta_n}_{\bm{k_n}})\approx S^{\alpha_0'\alpha'_j\theta_j}_3(\bm{k}_j-\bm{q}',\bm{q}',\bm{k}_j)\prod_{i\neq j}S_{k_i}^{\alpha'_i\theta_i}.
\end{eqnarray}
Combining these results, we  find that
\begin{eqnarray}
&&({\rho}^{\alpha'_0}_{\bm{k_j-q'}}\rho^{\alpha'_1}_{\bm{k_1}}\hdots{\rho}^{\alpha'_j}_{\bm{q'}}\hdots \rho^{\alpha'_n}_{\bm{k_n}}|f^{(n)}_{\{\mu_i\}}(\{k_i\})
=i\frac{1}{\sqrt{N}}\frac{k_BT}{m_{\mu_j}}\Big[(\bm{k}_j-\bm{q'})\cdot\bm{k}_j\delta_{\alpha'_0\mu_j}S_{\bm{q}'}^{\alpha'_j\mu_j}\prod_{i\neq j}S_{k_i}^{\alpha'_i\mu_i}
+\bm{q'}\cdot\bm{k}_j\delta_{\alpha'_j\mu_j}S_{|\bm{k}_j-\bm{q}'|}^{\alpha'_0\mu_j}\prod_{i\neq j}S_{k_i}^{\alpha'_i\mu_i}\Big]\nonumber
\\
&&-i\sum_{\{\theta_i\}\{\gamma_i\}}S^{\alpha_0'\alpha'_j\theta_j}_3(\bm{k}_j-\bm{q}',\bm{q}',\bm{k}_j)\prod_{i\neq j}S_{k_i}^{\alpha'_i\theta_i}\left(\prod_{i=1}^{n}S^{\theta_i\gamma_i}_{k_i}\right)^{-1}
\delta_{\gamma_j,\mu_j}\frac{k_BTx_{\gamma_j}k_j^2}{m_{\gamma_j}}\prod_{l\neq j}^{n}S^{\gamma_l\mu_l}_{k_l}
\nonumber
\\
&&=i\frac{1}{\sqrt{N}}\frac{k_BT}{m_{\mu_j}}\Big[(\bm{k}_j-\bm{q'})\cdot\bm{k}_j\delta_{\alpha'_0\mu_j}S_{\bm{q}'}^{\alpha'_j\mu_j}\prod_{i\neq j}S_{k_i}^{\alpha'_i\mu_i}
+\bm{q'}\cdot\bm{k}_j\delta_{\alpha'_j\mu_j}S_{|\bm{k}_j-\bm{q}'|}^{\alpha'_0\mu_j}\prod_{i\neq j}S_{k_i}^{\alpha'_i\mu_i}\nonumber
\\
&&-\sum_{\{\theta_i\}\{\gamma_i\}}
\sum_{\epsilon\sigma\eta}S^{\alpha_0'\epsilon}_{|\bm{k}_j-\bm{q'}|}S^{\alpha'_j\sigma}_{q'}S^{\theta_j\eta}_{k_j}[\delta_{\epsilon\sigma}\delta_{\sigma\eta}\delta_{\epsilon\eta}/x_{\epsilon}^2+{\rho}^2c_3^{\epsilon\sigma\eta}(\bm{k}_j-\bm{q}',\bm{q}')]\prod_{i\neq j}S_{k_i}^{\alpha'_i\theta_i}\left(\prod_{i=1}^{n}S^{\theta_i\gamma_i}_{k_i}\right)^{-1}
\delta_{\gamma_j,\mu_j}x_{\gamma_j}k_j^2\prod_{l\neq j}^{n}S^{\gamma_l\mu_l}_{k_l}\Big]
\nonumber
\\
&&=i\frac{1}{\sqrt{N}}\frac{k_BT}{m_{\mu_j}}\Big[(\bm{k}_j-\bm{q'})\cdot\bm{k}_j\delta_{\alpha'_0\mu_j}S_{\bm{q}'}^{\alpha'_j\mu_j}\prod_{i\neq j}S_{k_i}^{\alpha'_i\mu_i}
+\bm{q'}\cdot\bm{k}_j\delta_{\alpha'_j\mu_j}S_{|\bm{k}_j-\bm{q}'|}^{\alpha'_0\mu_j}\prod_{i\neq j}S_{k_i}^{\alpha'_i\mu_i}\nonumber
\\
&&-\sum_{\{\gamma_i\}}
\sum_{\epsilon\sigma\eta}S^{\alpha_0'\epsilon}_{|\bm{k}_j-\bm{q'}|}S^{\alpha'_j\sigma}_{q'}[\delta_{\epsilon\sigma}\delta_{\sigma\eta}\delta_{\epsilon\eta}/x_{\epsilon}^2+{\rho}^2c_3^{\epsilon\sigma\eta}(\bm{k}_j-\bm{q}',\bm{q}')]\prod_{i\neq j}\delta_{\alpha'_i\gamma_i}\delta_{\eta\gamma_j}
\delta_{\gamma_j\mu_j}x_{\gamma_j}k_j^2\prod_{l\neq j}^{n}S^{\gamma_l\mu_l}_{k_l}\Big]
\nonumber
\\
&&=i\frac{1}{\sqrt{N}}\frac{k_BT}{m_{\mu_j}}\Big[(\bm{k}_j-\bm{q'})\cdot\bm{k}_j\delta_{\alpha'_0\mu_j}S_{\bm{q}'}^{\alpha'_j\mu_j}\prod_{i\neq j}S_{k_i}^{\alpha'_i\mu_i}
+\bm{q'}\cdot\bm{k}_j\delta_{\alpha'_j\mu_j}S_{|\bm{k}_j-\bm{q}'|}^{\alpha'_0\mu_j}\prod_{i\neq j}S_{k_i}^{\alpha'_i\mu_i}\nonumber
\\
&&-
\sum_{\epsilon\sigma}S^{\alpha_0'\epsilon}_{|\bm{k}_j-\bm{q'}|}S^{\alpha'_j\sigma}_{q'}[\delta_{\epsilon\sigma}\delta_{\sigma\mu_j}\delta_{\epsilon\mu_j}/x_{\epsilon}^2+{\rho}^2c_3^{\epsilon\sigma\mu_j}(\bm{k}_j-\bm{q}',\bm{q}')]
x_{\mu_j}k_j^2\prod_{i\neq j}^{n}S^{\alpha'_i\mu_i}_{k_i}\Big]
\nonumber
\\
&&=i\frac{1}{\sqrt{N}}\frac{k_BT}{m_{\mu_j}}\Big[(\bm{k}_j-\bm{q'})\cdot\bm{k}_j\delta_{\alpha'_0\mu_j}S_{\bm{q}'}^{\alpha'_j\mu_j}\prod_{i\neq j}S_{k_i}^{\alpha'_i\mu_i}
+\bm{q'}\cdot\bm{k}_j\delta_{\alpha'_j\mu_j}S_{|\bm{k}_j-\bm{q}'|}^{\alpha'_0\mu_j}\prod_{i\neq j}S_{k_i}^{\alpha'_i\mu_i}\nonumber
\\
&&-
S^{\alpha_0'\mu_j}_{|\bm{k}_j-\bm{q'}|}S^{\alpha'_j\mu_j}_{q'}
k_j^2\prod_{i\neq j}^{n}S^{\alpha'_i\mu_i}_{k_i}/x_{\mu_j}
-
\sum_{\epsilon\sigma}S^{\alpha_0'\epsilon}_{|\bm{k}_j-\bm{q'}|}S^{\alpha'_j\sigma}_{q'}{\rho}^2c_3^{\epsilon\sigma\mu_j}(\bm{k}_j-\bm{q}',\bm{q}')
x_{\mu_j}k_j^2\prod_{i\neq j}^{n}S^{\alpha'_i\mu_i}_{k_i}\Big]
\nonumber
\\
&&=-i\frac{{\rho}}{\sqrt{N}}\frac{k_BT}{m_{\mu_j}}\Big[(\bm{k}_j-\bm{q'})\cdot\bm{k}_j\sum_{\theta}S^{\alpha_0'\theta}_{|\bm{k}_j-\bm{q'}|}S_{\bm{q}'}^{\alpha'_j\mu_j}\prod_{i\neq j}S_{k_i}^{\alpha'_i\mu_i}c^{\theta\mu_j}_{|\bm{k}_j-\bm{q'}|}
+\bm{q'}\cdot\bm{k}_j\sum_{\theta}S_{q'}^{\alpha'_j\theta}S_{|\bm{k}_j-\bm{q}'|}^{\alpha'_0\mu_j}\prod_{i\neq j}S_{k_i}^{\alpha'_i\mu_i}c^{\theta\mu_j}_{q'}\nonumber
\\
&&+
\sum_{\epsilon\sigma}S^{\alpha_0'\epsilon}_{|\bm{k}_j-\bm{q'}|}S^{\alpha'_j\sigma}_{q'}{\rho}^2c_3^{\epsilon\sigma\mu_j}(\bm{k}_j-\bm{q}',\bm{q}')
x_{\mu_j}k_j^2\prod_{i\neq j}^{n}S^{\alpha'_i\mu_i}_{k_i}\Big].
\end{eqnarray}

Invoking the normalization,  Eq.~(\ref{eq:TGMCT}), the projected fluctuating force can then be written as
\begin{eqnarray}
&&P_{j}^{(n)}|f^{(n)}_{\{\mu_i\}}(\{k_i\})\nonumber
\\
&&=\sum_{\bm{q'}\alpha'_0,\hdots,\alpha'_n}^{\bm{q},\alpha_0,\hdots,\alpha_n}|\rho^{\alpha_0}_{\bm{k_j-q}}\rho^{\alpha_1}_{\bm{k_1}}\hdots\rho^{\alpha_j}_{\bm{q}}\hdots \rho^{\alpha_n}_{\bm{k_n}})
T_{\bm{q'}\alpha'_0,\hdots,\alpha'_n}^{\bm{q},\alpha_0,\hdots,\alpha_n}(j)
(\rho^{\alpha'_0}_{\bm{k_j-q'}}\rho^{\alpha'_1}_{\bm{k_1}}\hdots\rho^{\alpha'_j}_{\bm{q'}}\hdots \rho^{\alpha'_n}_{\bm{k_n}}|f^{(n)}_{\{\mu_i\}}(\{k_i\})\nonumber
\\
&&=-i\frac{{\rho}}{\sqrt{N}}\frac{k_BT}{m_{\mu_j}}\sum_{\bm{q'}\alpha'_0,\hdots,\alpha'_n}^{\bm{q},\alpha_0,\hdots,\alpha_n}|\rho^{\alpha_0}_{\bm{k_j-q}}\rho^{\alpha_1}_{\bm{k_1}}\hdots\rho^{\alpha_j}_{\bm{q}}\hdots \rho^{\alpha_n}_{\bm{k_n}})
T_{\bm{q'}\alpha'_0,\hdots,\alpha'_n}^{\bm{q},\alpha_0,\hdots,\alpha_n}(j)
\nonumber
\\
&&\Big[(\bm{k}_j-\bm{q'})\cdot\bm{k}_j\sum_{\theta}S^{\alpha_0'\theta}_{|\bm{k}_j-\bm{q'}|}S_{\bm{q}'}^{\alpha'_j\mu_j}\prod_{i\neq j}S_{k_i}^{\alpha'_i\mu_i}c^{\theta\mu_j}_{|\bm{k}_j-\bm{q'}|}
+\bm{q'}\cdot\bm{k}_j\sum_{\theta}S_{q'}^{\alpha'_j\theta}S_{|\bm{k}_j-\bm{q}'|}^{\alpha'_0\mu_j}\prod_{i\neq j}S_{k_i}^{\alpha'_i\mu_i}c^{\theta\mu_j}_{q'}\nonumber
\\
&&+
\sum_{\epsilon\sigma}S^{\alpha_0'\epsilon}_{|\bm{k}_j-\bm{q'}|}S^{\alpha'_j\sigma}_{q'}{\rho}^2c_3^{\epsilon\sigma\mu_j}(\bm{k}_j-\bm{q}',\bm{q}')
x_{\mu_j}k_j^2\prod_{i\neq j}^{n}S^{\alpha'_i\mu_i}_{k_i}\Big]
\nonumber
\\
&&=-i\frac{{\rho}}{2\sqrt{N}}\frac{k_BT}{m_{\mu_j}}\sum^{\bm{q},\alpha_0,\hdots,\alpha_n}|\rho^{\alpha_0}_{\bm{k_j-q}}\rho^{\alpha_1}_{\bm{k_1}}\hdots\rho^{\alpha_j}_{\bm{q}}\hdots \rho^{\alpha_n}_{\bm{k_n}})
\nonumber
\\
&&\Big[(\bm{k}_j-\bm{q})\cdot\bm{k}_j\delta_{\alpha_j\mu_j}\prod_{i\neq j}\delta_{\alpha_i\mu_i}c^{\alpha_0\mu_j}_{|\bm{k}_j-\bm{q'}|}+\bm{q}\cdot\bm{k}_j\delta_{\alpha_0\mu_j}\prod_{i\neq j}\delta_{\alpha_i\mu_i}c^{\alpha_j\mu_j}_{q}\nonumber
+
{\rho}^2c_3^{\alpha_0\alpha_j\mu_j}(\bm{k}_j-\bm{q},\bm{q})
x_{\mu_j}k_j^2\prod_{i\neq j}^{n}\delta_{\alpha_i\mu_i}\Big]
\\
&&=-i\frac{{\rho}}{2\sqrt{N}}\frac{k_BT}{m_{\mu_j}}\sum^{\bm{q},\alpha_0,\alpha_j}|\rho^{\alpha_0}_{\bm{k_j-q}}\rho^{\mu_1}_{\bm{k_1}}\hdots\rho^{\alpha_j}_{\bm{q}}\hdots \rho^{\mu_n}_{\bm{k_n}})
\nonumber
\\
&&\Big[(\bm{k}_j-\bm{q})\cdot\bm{k}_j\delta_{\alpha_j\mu_j}c^{\alpha_0\mu_j}_{|\bm{k}_j-\bm{q}|}+\bm{q}\cdot\bm{k}_j\delta_{\alpha_0\mu_j}c^{\alpha_j\mu_j}_{q}\nonumber
+
{\rho}^2c_3^{\alpha_0\alpha_j\mu_j}(\bm{k}_j-\bm{q},\bm{q})
x_{\mu_j}k_j^2\Big]
\\
&&=-i\frac{{\rho}}{2\sqrt{N}}\frac{k_BT}{m_{\mu_j}}\sum^{\bm{q},\alpha_0,\alpha_j}|\rho^{\alpha_0}_{\bm{k_j-q}}\rho^{\mu_1}_{\bm{k_1}}\hdots\rho^{\alpha_j}_{\bm{q}}\hdots \rho^{\mu_n}_{\bm{k_n}})\mathcal{V}_{\alpha_j\alpha_0\mu_j}(\bm{q},\bm{k_j-q},\bm{k}_j),
\end{eqnarray}
such that the memory function simplifies to 
\begin{eqnarray}
&&K^{(n)}_{\{\mu_i\};\{\nu_i\}}(\{k_i\},\tau)=
(f^{(n)}_{\{\mu_i\}}(\{k_i\})P^{(n)}|e^{i\mathcal{L}t}P^{(n)}f^{(n)}_{\{\nu_i\}}(\{k_i\},\tau))\nonumber\\
&&=
(f^{(n)}_{\{\mu_i\}}(\{k_i\})\sum_lP^{(n)}_l|e^{i\mathcal{L}t}\sum_jP^{(n)}_jf^{(n)}_{\{\nu_i\}}(\{k_i\},\tau))\nonumber\\
&&=\frac{{\rho}^2}{4N}\sum_l\sum_j\sum_{\bm{q},\alpha_0,\alpha_l}\sum_{\bm{q'},\alpha'_0,\alpha'_j}\frac{k_BT}{m_{\mu_l}}\mathcal{V}_{\alpha_l\alpha_0\mu_l}(\bm{q},\bm{k_l-q},\bm{k}_l)
\nonumber\\
&&(\rho^{\alpha_0}_{\bm{k_l-q}}\rho^{\mu_1}_{\bm{k_1}}\hdots\rho^{\alpha_l}_{\bm{q}}\hdots \rho^{\mu_n}_{\bm{k_n}}|\rho^{\alpha'_0}_{\bm{k_j-q'}}(\tau)\rho^{\nu_1}_{\bm{k_1}}(\tau)\hdots\rho^{\alpha'_j}_{\bm{q'}}(\tau)\hdots \rho^{\nu_n}_{\bm{k_n}}(\tau))
\nonumber\\
&&
\mathcal{V}_{\alpha'_j\alpha'_0\nu_j}(\bm{q'},\bm{k_j-q'},\bm{k}_j)\frac{k_BT}{m_{\nu_j}},
\end{eqnarray}
with the same static vertices as in standard MCT [Eq.\ (\ref{eq:vertex})]. 
For convenience 
we will only focus on the diagonal terms of the dynamical multi-component density correlation functions, i.e.\
\begin{eqnarray}
&&(\rho^{\alpha_0}_{\bm{k_l-q}}\rho^{\mu_1}_{\bm{k_1}}\hdots\rho^{\alpha_l}_{\bm{q}}\hdots \rho^{\mu_n}_{\bm{k_n}}|\rho^{\alpha'_0}_{\bm{k_j-q'}}(\tau)\rho^{\nu_1}_{\bm{k_1}}(\tau)\hdots\rho^{\alpha'_j}_{\bm{q'}}(\tau)\hdots \rho^{\nu_n}_{\bm{k_n}}(\tau))
\nonumber\\
&&
\approx
(\rho^{\alpha_0}_{\bm{k_l-q}}\rho^{\mu_1}_{\bm{k_1}}\hdots\rho^{\alpha_l}_{\bm{q}}\hdots \rho^{\mu_n}_{\bm{k_n}}|\rho^{\alpha'_0}_{\bm{k}_l-\bm{q}}(\tau)\rho^{\nu_1}_{\bm{k_1}}(\tau)\hdots\rho^{\alpha'_l}_{\bm{q}}(\tau)\hdots \rho^{\nu_n}_{\bm{k_n}}(\tau))\delta_{\bm{qq'}}\delta_{lj}
\nonumber\\
&&
+(\rho^{\alpha_0}_{\bm{k_l-q}}\rho^{\mu_1}_{\bm{k_1}}\hdots\rho^{\alpha_l}_{\bm{q}}\hdots \rho^{\mu_n}_{\bm{k_n}}|\rho^{\alpha'_0}_{\bm{q}}(\tau)\rho^{\nu_1}_{\bm{k_1}}(\tau)\hdots\rho^{\alpha'_l}_{\bm{k_l-q}}(\tau)\hdots \rho^{\nu_n}_{\bm{k_n}}(\tau))\delta_{\bm{q, k_j-q'}}\delta_{lj},
\end{eqnarray}
which implies that the memory function reduces to 
\begin{eqnarray}
&&K^{(n)}_{\{\mu_i\};\{\nu_i\}}(\{k_i\},\tau)
\nonumber\\
&&=\frac{{\rho}^2}{4N}\sum_j\sum_{\bm{q}}\sum_{\alpha_0,\alpha_j,\alpha'_0,\alpha'_j}\frac{k_BT}{m_{\mu_j}}\mathcal{V}_{\alpha_j\alpha_0\mu_j}(\bm{q},\bm{k_j-q},\bm{k}_j)
\nonumber\\
&&(\rho^{\alpha_0}_{\bm{k_j-q}}\rho^{\mu_1}_{\bm{k_1}}\hdots\rho^{\alpha_j}_{\bm{q}}\hdots \rho^{\mu_n}_{\bm{k_n}}|\rho^{\alpha'_0}_{\bm{k_j-q}}(\tau)\rho^{\nu_1}_{\bm{k_1}}(\tau)\hdots\rho^{\alpha'_j}_{\bm{q}}(\tau)\hdots \rho^{\nu_n}_{\bm{k_n}}(\tau))
\nonumber\\
&&
\mathcal{V}_{\alpha'_j\alpha'_0\nu_j}(\bm{q},\bm{k_j-q},\bm{k}_j)\frac{k_BT}{m_{\nu_j}}
\nonumber\\
&&+\frac{{\rho}^2}{4N}\sum_j\sum_{\bm{q}}\sum_{\alpha_0,\alpha_j,\alpha'_0,\alpha'_j}\frac{k_BT}{m_{\mu_j}}\mathcal{V}_{\alpha_j\alpha_0\mu_j}(\bm{q},\bm{k_j-q},\bm{k}_j)
\nonumber\\
&&(\rho^{\alpha_0}_{\bm{k_j-q}}\rho^{\mu_1}_{\bm{k_1}}\hdots\rho^{\alpha_j}_{\bm{q}}\hdots \rho^{\mu_n}_{\bm{k_n}}|\rho^{\alpha'_0}_{\bm{q}}(\tau)\rho^{\nu_1}_{\bm{k_1}}(\tau)\hdots\rho^{\alpha'_j}_{\bm{k_j-q}}(\tau)\hdots \rho^{\nu_n}_{\bm{k_n}}(\tau))
\nonumber\\
&&
\mathcal{V}_{\alpha'_j\alpha'_0\nu_j}(\bm{k_j-q},\bm{q},\bm{k}_j)\frac{k_BT}{m_{\nu_j}}
\\
&&=\frac{{\rho}^2}{2N}\sum_j\sum_{\bm{q}}\sum_{\alpha_0,\alpha_j,\alpha'_0,\alpha'_j}\frac{k_BT}{m_{\mu_j}}\mathcal{V}_{\alpha_j\alpha_0\mu_j}(\bm{q},\bm{k_j-q},\bm{k}_j)
\nonumber\\
&&(\rho^{\alpha_0}_{\bm{k_j-q}}\rho^{\mu_1}_{\bm{k_1}}\hdots\rho^{\alpha_j}_{\bm{q}}\hdots \rho^{\mu_n}_{\bm{k_n}}|\rho^{\alpha'_0}_{\bm{k_j-q}}(\tau)\rho^{\nu_1}_{\bm{k_1}}(\tau)\hdots\rho^{\alpha'_j}_{\bm{q}}(\tau)\hdots \rho^{\nu_n}_{\bm{k_n}}(\tau))
\nonumber\\
&&
\mathcal{V}_{\alpha'_j\alpha'_0\nu_j}(\bm{q},\bm{k_j-q},\bm{k}_j)\frac{k_BT}{m_{\nu_j}}
\nonumber\\
&&=\frac{{\rho}}{2}\int\frac{d\bm{q}}{(2\pi)^3}\sum_j\sum_{\alpha_0,\alpha_j,\alpha'_0,\alpha'_j}\frac{k_BT}{m_{\mu_j}}\mathcal{V}_{\alpha_j\alpha_0\mu_j}(\bm{q},\bm{k_j-q},\bm{k}_j)
\nonumber\\
&&(\rho^{\alpha_0}_{\bm{k_j-q}}\rho^{\mu_1}_{\bm{k_1}}\hdots\rho^{\alpha_j}_{\bm{q}}\hdots \rho^{\mu_n}_{\bm{k_n}}|\rho^{\alpha'_0}_{\bm{k_j-q}}(\tau)\rho^{\nu_1}_{\bm{k_1}}(\tau)\hdots\rho^{\alpha'_j}_{\bm{q}}(\tau)\hdots \rho^{\nu_n}_{\bm{k_n}}(\tau))
\nonumber\\
&&
\mathcal{V}_{\alpha'_j\alpha'_0\nu_j}(\bm{q},\bm{k_j-q},\bm{k}_j)\frac{k_BT}{m_{\nu_j}}
\nonumber\\
&&=\frac{{\rho}}{2}\int\frac{d\bm{q}}{(2\pi)^3}\sum_j\sum_{\alpha_0,\alpha_j,\alpha'_0,\alpha'_j}\frac{k_BT}{m_{\mu_j}}\mathcal{V}_{\alpha_j\alpha_0\mu_j}(\bm{q},\bm{k_j-q},\bm{k}_j)
\nonumber\\
&&F^{(n+1)}_{\alpha_0\mu_1\hdots\alpha_j\hdots\mu_n;\alpha'_0\nu_1\hdots\alpha'_j\hdots\nu_n}(\bm{k_j-q},\bm{k_1},\hdots,{\bm{q}},\hdots,\bm{k_n},\tau)
\nonumber\\
&&
\mathcal{V}_{\alpha'_j\alpha'_0\nu_j}(\bm{q},\bm{k_j-q},\bm{k}_j)\frac{k_BT}{m_{\nu_j}}.
\end{eqnarray}
Note that in going from the second to the third equality we have made use of  $\mathcal{V}_{\alpha_0\alpha_j\mu_j}(\bm{k_j-q},\bm{q},\bm{k}_j)=\mathcal{V}_{\alpha_j\alpha_0\mu_j}(\bm{q},\bm{k_j-q},\bm{k}_j)$.
Finally, we can rewrite the memory kernel in its more compact final form, 
\begin{eqnarray}
&&K^{(n)}_{\{\alpha_i\};\{\beta_i\}}(\{k_i\},\tau)
\nonumber\\
&&=\frac{\rho}{2}\sum_{\mu'\nu'}\sum_{\mu\nu}\int \frac{d\mathbf{q}}{(2\pi)^3}
\sum_{j=1}^{n}\frac{k_BT}{m_{\alpha_j}}\mathcal{V}_{\mu'\nu'\alpha_j}(\bm{q,k_j-q,k_j})F^{(n+1)}_{\mu',\nu',\{\alpha_i\}/\alpha_j;\mu,\nu,\{\beta_i\}/\beta_j}(\bm{q,k_j-q},\{k_i\}/k_j,\tau) 
\nonumber\\
&&
\mathcal{V}_{\mu\nu\beta_j}(\bm{q,k_j-q,k_j})\frac{k_BT}{m_{\beta_j}}~,
\label{eq:Kn}
\end{eqnarray}
which is the expression presented in the main text [Eq.~(6)]. As a final remark, note that the factor $c_3$ in the vertex [Eq.~(\ref{eq:vertex})] is neglected in the main work (also known as the convolution approximation), since $c_3$ generally does not make a dominant contribution for fragile systems~\cite{sciortino2001debye}.

In order to solve the GMCT equations, we have to close the hierarchy at a finite level $n_\mathrm{max}$. In the spirit of mean-field models, we choose to approximate the last level
$F^{(n_\mathrm{max})}$
in the memory kernel in terms of the lower level correlation functions $F^{(n_\mathrm{max}-1)}$ and $F^{(1)}$,
\begin{eqnarray}
\label{eq:Fn-1}
&&F^{(n_\mathrm{max})}_{\mu',\nu',\{\alpha_i\}/\alpha_j;\mu,\nu,\{\beta_i\}/\beta_j}(\bm{q,k_j-q},\{k_i\}/k_j,\tau)
\nonumber\\
&&\approx\frac{1}{n_\mathrm{max}-2}\sum_{l\neq j} F^{(n_\mathrm{max}-1)}_{\mu',\nu',\{\alpha_i\}/\{\alpha_j,\alpha_l\};\mu,\nu,\{\beta_i\}/\{\beta_j,\beta_l\}}(\bm{q,k_j-q},\{k_i\}/\{k_j,k_l\},\tau)F^{(1)}_{\alpha_l\beta_l}(k_l,\tau).
\end{eqnarray}
This leads to the following closure for the memory function at one level lower, i.e.\ at level $n_\mathrm{max}-1$,
\begin{eqnarray}
\label{eq:Kn-1}
K^{(n_\mathrm{max}-1)}_{\{\alpha_i\};\{\beta_i\}}(\{k_i\},t)
\approx
\frac{1}{n_\mathrm{max}-2}\sum_{j=1}^{n_\mathrm{max}-1}
K^{(n_\mathrm{max}-2)}_{\{\alpha_i\}/\alpha_j;\{\beta_i\}/\beta_j}(\{k_i\}/k_j,t)
 F^{(1)}_{\alpha_j;\beta_j}(k_j,t).
 \end{eqnarray}
Note that this choice of closure allows us to directly obtain the memory kernel $K^{(n_\mathrm{max}-1)}$ from  $K^{(n_\mathrm{max}-2)}$, which 
is computationally significantly faster than first estimating $F^{(n_\mathrm{max})}$ from Eq.~(\ref{eq:Fn-1}) and subsequently calculating $K^{(n_\mathrm{max}-1)}$ from Eq.~({\ref{eq:Kn})}.

\subsection{Numerical details}
\begin{figure}
    \centering
    \includegraphics[width=0.5\textwidth]{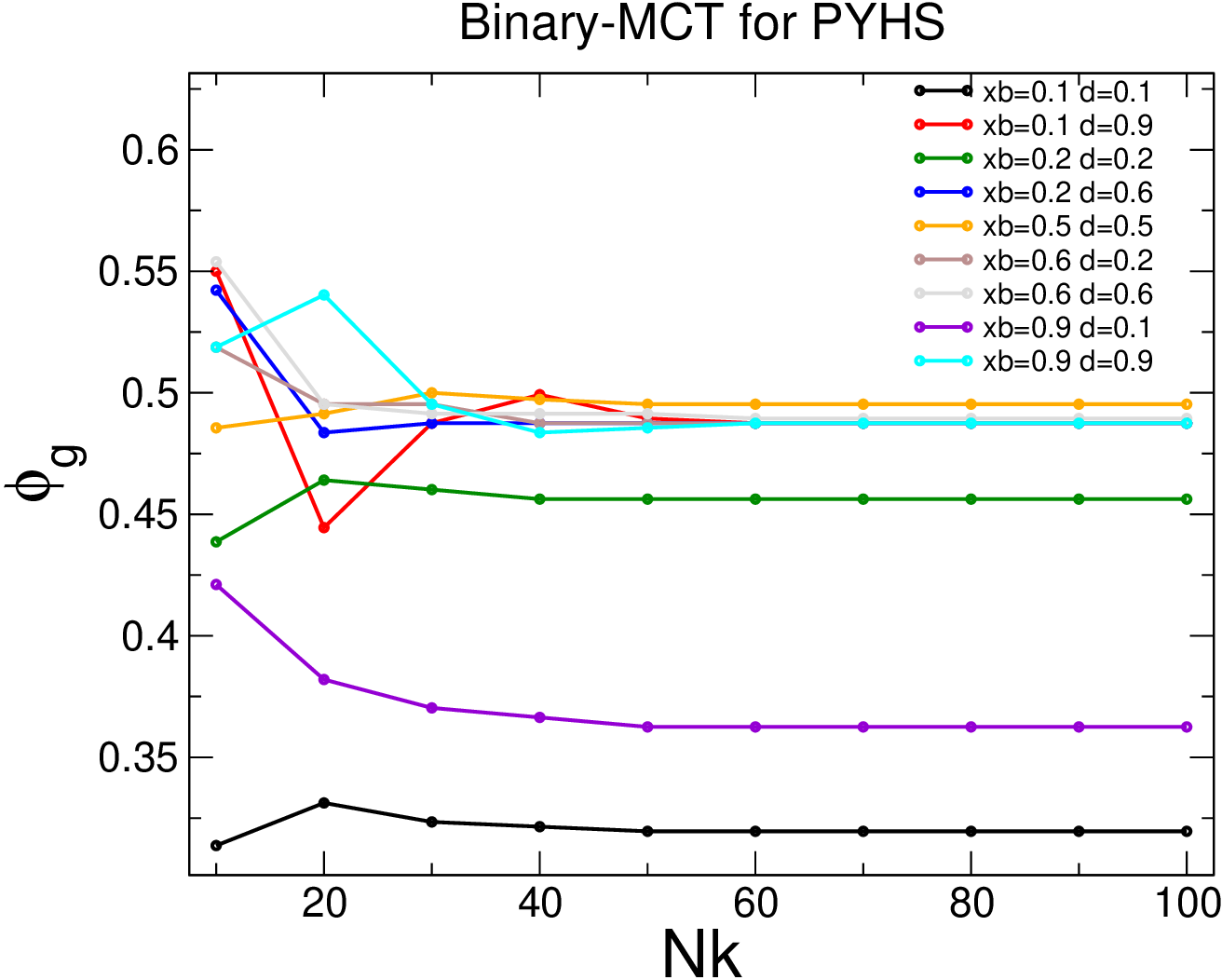}
    \caption{Predicted glass-transition packing fraction $\varphi_g$ as a function of the size of the wavenumber grid, $N_k$, for binary MCT applied to a binary mixture of hard spheres described by the Percus-Yevick closure. We report different values of the small-particle concentration $x_b$ and the particle size ratio $d$. Finite size effects appear only around $N_k\sim60$. We have assumed that a similar $N_k$ convergence holds for binary GMCT applied to LJ and WCA mixtures; in the main manuscript we have used $N_k=70$.}
    \label{fig:nk}
\end{figure}
In the main manuscript we report that we used a grid of $N_k=70$ wavenumbers to solve multi-component GMCT.
The reason we want to use a small $N_k$ is that the complexity of multi-component GMCT scales as $~N_k^n$, where $n$ is the GMCT order.
To justify the choice of $N_k=70$ we report in Fig.~\ref{fig:nk} the effect of the variation of $N_k$ over the critical packing fraction $\phi_g$ above which a binary mixture of hard spheres, described using the Percus-Yevick closure, is a glass~\cite{Gotze2003}.
Here we have considered different values of the concentration of small particles $x_b$ and the size ratio $d=r_b/r_a$ where $r_\alpha$ is the radius of the hard sphere of species $\alpha$. Overall we see that for any binary mixture composition $N_k=70$ produces the asymptotic value of $\phi_g$, and finite size effects are only visible below $N_k~60$.
As a result we use $N_k=70$ to solve higher order GMCT in the main manuscript.
Lastly, in Table I we report the fit parameters used to produce Fig.~4 of the main manuscript.

\setlength{\tabcolsep}{2em}
\begin{table}
\centering
    \begin{tabular}{|c|c|c|c|c|c|}
      \hline  
      \textbf{Model} & \textbf{Theory} & $\mathbf{\rho}$ & $\mathbf{T_0}$ & $\mathbf{\gamma}$ & $\mathbf{A_0}$ \\
      \hline
       KABLJ & simulations & 1.2 & 2.7059 & 0.5519 & 0.0000 \\
       KABLJ & (G)MCT $n_\mathrm{max}=2$ & 1.2 & 1.3500 & 0.5773 & 0.6481 \\
       KABLJ & GMCT $n_\mathrm{max}=3$ & 1.2 & 1.7101 & 0.5411 & 0.8161 \\
       KABLJ & GMCT $n_\mathrm{max}=4$ & 1.2 & 2.4964 & 0.7768 & 0.7585 \\
       WCA & simulations & 1.2 & 4.9973 & 0.5895 & 0.0000 \\
       WCA & (G)MCT $n_\mathrm{max}=2$ & 1.2 & 1.6384 & 0.5416 & 0.7391 \\
       WCA & GMCT $n_\mathrm{max}=3$ & 1.2 & 2.2052 & 0.5580 & 0.8076 \\
       WCA & GMCT $n_\mathrm{max}=4$ & 1.2 & 5.7106 & 1.0117 & 0.8507 \\
       KABLJ & simulations & 1.4 & 2.0890 & 1.7257 & 0.5583 \\
       KABLJ & (G)MCT $n_\mathrm{max}=2$ & 1.4 & 0.6988 & 1.2428 & 0.5555 \\
       KABLJ & GMCT $n_\mathrm{max}=3$ & 1.4 & 0.8513 & 0.8260 & 0.8668 \\
       KABLJ & GMCT $n_\mathrm{max}=4$ & 1.4 & 1.2207 & 1.2159 & 1.0930 \\
       WCA & simulations & 1.4 & 3.0898 & 1.9598 & 0.6124 \\
       WCA & (G)MCT $n_\mathrm{max}=2$ & 1.4 & 0.7349 & 1.1045 & 0.4735 \\
       WCA & GMCT $n_\mathrm{max}=3$ & 1.4 & 0.9077 & 0.8043 & 0.8751 \\
       WCA & GMCT $n_\mathrm{max}=4$ & 1.4 & 1.4005 & 1.2782 & 1.1261 \\
      \hline
    \end{tabular}
    \caption{The table contains the fitting parameters used to produce Fig.~4 (main manuscript).}
    \label{tab:table1}
\end{table}


\end{document}